\documentclass{aa}
\usepackage{natbib}
\bibpunct{(}{)}{;}{a}{}{,}

\usepackage[dvips]{graphicx}
\usepackage{times}
\usepackage{rotating}
\usepackage{longtable}
\newcommand\ignore[1]{} 

\begin{document}
\title{Hipparcos red stars in the $HpV_{T2}$ and $V\!I_{C}$ systems
\thanks{Based on observations from the Hipparcos astrometric satellite 
operated by the European Space Agency (ESA 1997)}\fnmsep\thanks{
Table 7 is available only in electronic form at the Centre de Donn\'{e}es
Astronomiques de Strasbourg (CDS), France via anonymous ftp to 
cdsarc.u-strasbg.fr (130.79.128.5) or at
http://cdsweb.u-strasbg.fr/cgi-bin/qcat?J/A+A/(vol)/(page)}}
\titlerunning{Hipparcos red stars} 
\author{I.~Platais\inst{1,2}
\and
D.~Pourbaix\inst{1}\fnmsep\thanks{Research Associate, F.N.R.S., Belgium} 
\and
A.~Jorissen\inst{1}\fnmsep$^{\star\star\star}$ 
\and
V.~V.~Makarov\inst{2,3}
\and
L.~N.~Berdnikov\inst{4}
\and
N~.N.~Samus\inst{5,4}
\and
T.~Lloyd Evans\inst{6}
\and
T.~Lebzelter\inst{7}
\and
J.~Sperauskas\inst{8}
}
\institute{
Institut d'Astronomie et d'Astrophysique, Universit\'e Libre de Bruxelles, 
C.P.~226, Boulevard du Triomphe, B-1050 Bruxelles, Belgium 
\and 
Universities Space Research Association, Division of Astronomy and Space
Physics, 300 D Street SW, Washington, D.C. 20024, U.S.A.
\and
U.S. Naval Observatory, 3450 Massachusetts Ave., NW, Washington D.C.
20392-5420, U.S.A.
\and
Sternberg Astronomical Institute and Isaac Newton Institute of Chile,
Moscow Branch, 13 Universitetskij Prosp., Moscow 119992, Russia
\and
Institute of Astronomy, Russian Academy of Sciences, 48 Pyatnitskaya Str.,
Moscow 119017, Russia
\and
School of Physics and Astronomy, University of St Andrews, North Haugh, St Andrews, Fife, Scotland KY16 9SS
\and
Institut f\"{u}r Astronomie, Universit\"{a}t Wien, T\"{u}rkenschanzstr. 17,
1180 Vienna, Austria
\and
Vilnius University Observatory, Ciurlionio 29, Vilnius 2009, Lithuania
}
\date{Received date; accepted date} 
\offprints{pourbaix@astro.ulb.ac.be}
\abstract{
For Hipparcos M, S, and C spectral type stars, we provide calibrated 
instantaneous (epoch) Cousins $V\!-\!I$ color indices
using newly derived $HpV_{T2}$ photometry. 
Three new sets of ground-based Cousins $V\!I$ data have been
obtained for more than 170 carbon and red M giants. These datasets
in combination with the published sources of $V\!I$ photometry served
to obtain the calibration curves linking Hipparcos/Tycho $Hp-V_{T2}$ with
the Cousins $V\!-\!I$ index. In total, 321 carbon stars and 4464 M- and
S-type stars have new $V\!-\!I$ indices. The standard error of the
mean $V\!-\!I$ is about 0.1 mag or better down to $Hp\approx9$ 
although it deteriorates rapidly at fainter magnitudes. 
These $V\!-\!I$ indices can be used 
to verify the published Hipparcos $V\!-\!I$ color indices. Thus, we
have identified a handful of new cases where, instead of the real
target, a random field star has been observed. 
A considerable fraction of the DMSA/C and DMSA/V 
solutions for red stars appear not to be warranted. Most likely such
spurious solutions may originate from usage of a heavily biased
color in the astrometric processing.
}
\maketitle

%
%------------------------------------------------------------------------------
\section{Introduction}
%------------------------------------------------------------------------------
%

The Hipparcos Catalogue \citep{esa97} includes two sets of Cousins $V\!-\!I$
color indices -- a functional $V\!-\!I$ (entry H75 in the main Hipparcos
Catalogue) and
a best available $V\!-\!I$ at the time of the Catalogue's release (entry H40).  
This color index is an important temperature indicator for late-type
stars \citep{dumm98,bessell98}. 
Since only 2989 Hipparcos stars are listed
as having direct measurements of the Cousins $V\!-\!I$ index, nineteen 
different
methods of variable accuracy were used to obtain $V\!-\!I$ photometry 
(see ESA 1997, Sect. 1.3, Appendix 5). In numerous cases the reductions  
of Hipparcos $V\!-\!I$ photometry relied heavily upon the satellite's 
star mapper photometry -- the Tycho $B_T-V_{T}$ color indices. However,
the Tycho photometric system alone is not well-suited for the studies of 
fainter red stars.  A combination of intrinsically low fluxes from these stars 
in the $B_T$ bandpass and a short crossing time ($\sim\!22$ ms) 
of the star mapper's four vertical slits resulted in low S/N ratios. 
This, in combination with the residual bias that was not fully
corrected by the de-censoring analysis \citep{halbwachs97} 
in deriving the Tycho photometry for faint stars, diminishes 
the reliability of much of the published Hipparcos 
$V\!-\!I$ indices for stars with $V\!-\!I\ga1.5$. As demonstrated by
\citet{koen02}, the listed Hipparcos $V\!-\!I$ photometry of red 
stars shows a disappointingly 
large scatter with respect to the ground-based photoelectric $V\!-\!I$
measurements. In extreme cases the disagreement can reach
up to 2-3 magnitudes.

Our interest in the $V\!-\!I$ photometry of red stars is primarily motivated
by the potential effect of incorrect $V\!-\!I$ color indices on the chromaticity
corrections in Hipparcos astrometry. On average, a one magnitude offset
in the $V\!-\!I$ value could introduce a $\sim\!1$ mas bias in the 
star's position
(abscissa) along the scan direction. Besides grossly incorrect $V\!-\!I$
indices for some red stars \citep{koen02}, there is a systematic color
bias related to neglecting in the Hipparcos reductions the intrinsic color
variation in large amplitude variables such as Miras.

In retrospect, the Hipparcos $V\!-\!I$ photometry would have gained
considerably from the parallel-in-time ground-based $V\!-\!I$ observations
of stars with extreme colors and/or considerable color variability.
For a number of reasons, most importantly, a prorogated decision to choose 
the $V\!-\!I$ index, this opportunity was lost. Is it possible to improve
the Hipparcos $V\!-\!I$ photometry now? Here we attempt to answer
this question. It appears that high-grade
 $V\!-\!I$ photometry for red stars
is possible down to $V\approx8$ and may even be used to obtain an estimate
of effective temperatures. In general, the re-calibrated $V\!-\!I$ photometry 
is useful in identifying some difficult cases in the Hipparcos Catalogue,
such as red and variable stars in binary systems.
Throughout the paper we refer to 
Cousins $V\!-\!I$ color indices, unless it is explicitly stated otherwise.

\section{Ground-based Cousins $V\!I$ photometry}\label{groundVI}

The advantages of the broadband Cousins $V\!RI$ photometric system such as high
internal precision and maintaining this precision over the whole range 
of spectral types are discussed by \citet{bessell79}. This system emerged 
with the advent of Ga-As photocathode photomultipliers in the early
1970s.  There are two issues which should be considered in the broadband 
photometry of red stars. First, the majority of cool red stars are
variable and no standard stars are available redder than $V\!-\!I\approx3$.
Second, the presence of numerous molecular bands in the spectra of
red stars requires stable and easily reproducible bandpasses in order
to avoid possible 
nonlinear transformations from the instrumental to the standard system. 
In other words, to exclude the transformation uncertainties, such stars
must be observed in the natural Cousins $V\!RI$ system, i.e. using the 
same filters and detector. Examination of the published sources of 
Cousins $V\!RI$ photometry indicates that many extremely red Hipparcos
stars actually lack this photometry.
Therefore, we have obtained new sets of $U\!BV\!RI$ photometry
of the Southern carbon stars and $BV\!I$ photometry for the reddest
M and C spectral-type stars.   

\subsection{Carbon star photometry at SAAO}

The observations of 85 carbon stars including a few hydrogen-deficient 
(Hd) stars were made in 1984 and 1987
using the single channel Modular Photometer on the
0.5m reflector at the Sutherland station of the SAAO. The photometer uses a
Hamamatsu R943-02 GaAs photomultiplier and a filter set which reproduces the
Johnson $U\!BV$ and Cousins $(RI)_{C}$ photometric systems, 
with a need for only very small linear and
non-linear terms in transformations onto the standard system. 
The observations were made with frequent
reference to the E-region standard stars of \citet{menzies89}. 
The results of $U\!BV\!(RI)_{C}$ photometry are provided in 
Table~\ref{tab:evans}. The CGCS numbers are those in \citet{stephenson89}. 
The last column indicates the total number of observations, usually
obtained over 2-3 nights. The standard error of individual observations is
about 0.01 mag. It was necessary, however, to extrapolate the
color system as some of the stars here are redder than any standard star in
one color or another and in any case these are carbon stars (or helium
stars in the case of Hd stars) whose colors differ
systematically from the oxygen-rich M spectral type standard stars.
In addition, most of our programme stars are variable to some degree.
All cases with apparent variability or uncertain photometry are marked by 
(v) or (:), accordingly. Since in
the $UBVRI$ photoelectric photometry the aperture size varied from 
$20\arcsec$ to $40\arcsec$, a nearby optical component, marked in
Table~\ref{tab:evans}, may affect the accuracy of our photometry. 
\begin{figure}[htb]
\resizebox{\hsize}{!}{\includegraphics{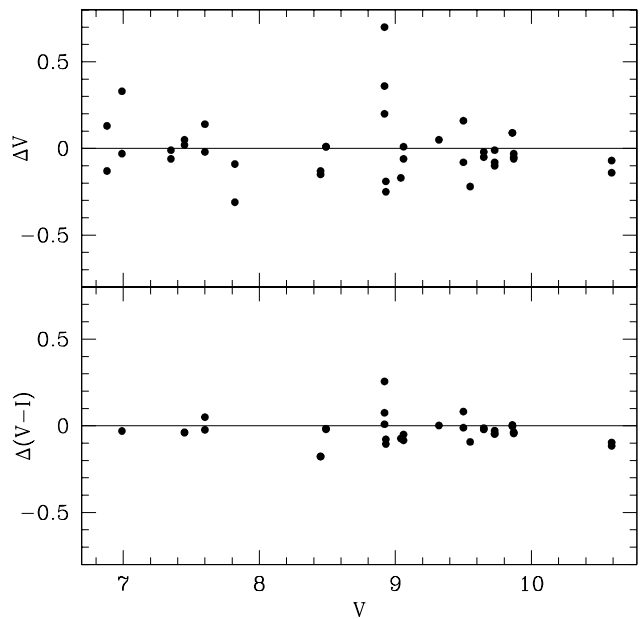}}
\caption[]{\label{fig:walk} Differences between the $V$-magnitude
(top panel) and $V\!-\!I$ color index (bottom panel) from 
Table~\ref{tab:evans} and that of \citet{walker79}. 
A relatively large spread visible in the top panel is mainly due to 
the variability.
}
\end{figure}
The generally good agreement (Fig.~\ref{fig:walk})
with the data of \citet{walker79}, whose observations were made with separate
blue and red sensitive photomultipliers and a different filter set, give
added confidence to the results. 
 
\begin{table*}[htb]
\caption[]{\label{tab:evans}SAAO photometry of selected carbon stars}
\setlength{\tabcolsep}{1mm}
\begin{tabular}{rrrrllllr|rrrrllllr}\hline
CGCS & HIP & GCVS & $V$\phantom{V}  & $B\!-\!V$ & 
$U\!-\!B$ & $R\!-\!I$ & $V\!-\!I$ & $n$ &
CGCS & HIP & GCVS & $V$\phantom{V}  & $B\!-\!V$ & 
$U\!-\!B$ & $R\!-\!I$ & $V\!-\!I$ & $n$ \\ \hline
 177\phantom{v} &         &     AM Scl & 12.33\phantom{v} &  2.16\phantom{v} &  3.42: &            1.02\phantom{v} &  1.85\phantom{v} &  3 & 3810\phantom{v} &         &            & 10.39\phantom{v} &  1.41\phantom{v} &  0.90\phantom{v} &  0.83\phantom{v} &  1.63\phantom{v} &  5\\
 196\phantom{v} &    5809 &            & 10.02\phantom{v} &  1.33\phantom{v} &  1.34\phantom{v} &  0.59\phantom{v} &  1.09\phantom{v} &  5 & 3813\phantom{v} &         &            & 11.02v &            2.39\phantom{v} &  2.61\phantom{v} &  1.30\phantom{v} &  2.45\phantom{v} &  4\\
 258\phantom{v} &         &            & 10.19\phantom{v} &  1.31\phantom{v} &  1.16\phantom{v} &  0.61\phantom{v} &  1.12\phantom{v} &  6 & 3842\phantom{v} &   85750 &            &  9.37\phantom{v} &  1.88\phantom{v} &  2.02\phantom{v} &  0.88\phantom{v} &  1.66\phantom{v} &  5\\
 327\phantom{v} &   10472 &      V Ari &  8.71\phantom{v} &  2.19\phantom{v} &  2.45\phantom{v} &  1.16\phantom{v} &  2.15\phantom{v} &  6 & 3855$^{\mathrm{a}}$ &         &            & 11.20\phantom{v} &  1.30\phantom{v} &  0.91\phantom{v} &  0.64\phantom{v} &  1.19\phantom{v} &  2\\
 378\phantom{v} &   12028 &            &  8.16\phantom{v} &  1.24\phantom{v} &  0.77\phantom{v} &  0.60\phantom{v} &  1.09\phantom{v} &  4 & 3864$^{\mathrm{a}}$ &         &   V450 Sco & 10.30v &            2.38\phantom{v} &  3.56: &            1.50\phantom{v} &  2.82\phantom{v} &  4\\
 576\phantom{v} &   17933 &            &  8.30\phantom{v} &  1.65\phantom{v} &  1.72\phantom{v} &  0.75\phantom{v} &  1.41\phantom{v} &  6 & 3938\phantom{v} &   88584 &      W CrA &  9.95\phantom{v} &  1.89\phantom{v} &  1.83\phantom{v} &  0.99\phantom{v} &  1.81\phantom{v} &  4\\
 639\phantom{v} &   19269 &            & 10.66\phantom{v} &  1.23\phantom{v} &  0.68\phantom{v} &  0.72\phantom{v} &  1.43\phantom{v} &  4 & 3939\phantom{v} &         &  V1783 Sgr & 10.53\phantom{v} &  1.54\phantom{v} &  1.01\phantom{v} &  0.94\phantom{v} &  1.74\phantom{v} &  3\\
 725\phantom{v} &   21051 &            &  8.91\phantom{v} &  1.14\phantom{v} &  1.12\phantom{v} &  0.55\phantom{v} &  1.04\phantom{v} &  6 & 3957\phantom{v} &   88887 &            &  9.80\phantom{v} &  1.52\phantom{v} &  1.18\phantom{v} &  0.96\phantom{v} &  1.87\phantom{v} &  4\\
1380\phantom{v} &   31725 &            &  9.37\phantom{v} &  1.37\phantom{v} &  1.28\phantom{v} &  0.58\phantom{v} &  1.07\phantom{v} &  3 & 3958\phantom{v} &         &            & 10.45\phantom{v} &  1.34\phantom{v} &  0.95\phantom{v} &  0.66\phantom{v} &  1.23\phantom{v} &  6\\
1460\phantom{v} &   33042 &     KY CMa & 10.75\phantom{v} &  2.73\phantom{v} &  4.00: &            1.33\phantom{v} &  2.42\phantom{v} &  4 & 3966$^{\mathrm{a}}$ &         &            & 11.06v &            1.95\phantom{v} &  1.76\phantom{v} &  1.05\phantom{v} &  1.95\phantom{v} &  4\\
1489\phantom{v} &   33550 &     RV Mon &  6.88\phantom{v} &  2.65\phantom{v} &  7.16: &                \phantom{v} &      \phantom{v} &  3 & 3992\phantom{v} &   89783 &     FO Ser &  8.42\phantom{v} &  1.85\phantom{v} &  1.85\phantom{v} &  1.15\phantom{v} &  2.21\phantom{v} &  3\\
1507\phantom{v} &   33794 &   V614 Mon &  7.32v &            1.76\phantom{v} &  2.14: &            1.13\phantom{v} &      \phantom{v} &  4 & 4021\phantom{v} &   90694 &            &  9.90\phantom{v} &  1.39\phantom{v} &  0.92\phantom{v} &  0.81\phantom{v} &  1.60\phantom{v} &  3\\
1659\phantom{v} &   35549 &     MY CMa & 10.63\phantom{v} &  2.44\phantom{v} &  3.08: &            1.36\phantom{v} &  2.55\phantom{v} &  3 & 4042\phantom{v} &         &            & 11.13\phantom{v} &  2.08\phantom{v} &  2.01\phantom{v} &  1.05\phantom{v} &  1.90\phantom{v} &  3\\
1790\phantom{v} &         &            &  9.58\phantom{v} &  1.85\phantom{v} &  2.15\phantom{v} &  1.11\phantom{v} &  2.07\phantom{v} &  3 & 4070\phantom{v} &         &            &  9.33\phantom{v} &  1.29\phantom{v} &  1.00\phantom{v} &  0.58\phantom{v} &  1.06\phantom{v} &  3\\
1871$^{\mathrm{a}}$ &         &            & 10.16\phantom{v} &  1.23\phantom{v} &  0.82\phantom{v} &  0.56\phantom{v} &  1.02\phantom{v} &  3 & 4086\phantom{v} &   91929 &     RV Sct & 10.02\phantom{v} &  2.35\phantom{v} &  2.52\phantom{v} &  1.54\phantom{v} &  2.92\phantom{v} &  3\\
1968\phantom{v} &   38787 &   V406 Pup &  7.62v &            3.20: &            4.60: &            1.40\phantom{v} &      \phantom{v} &  4 & 4094\phantom{v} &   92115 &            &  9.49\phantom{v} &  0.83\phantom{v} &  0.47\phantom{v} &  0.42\phantom{v} &  0.67\phantom{v} &  3\\
2153\phantom{v} &   40805 &   V433 Pup &  9.54v &            1.67\phantom{v} &  1.66\phantom{v} &  1.07\phantom{v} &  2.05\phantom{v} &  3 & 4145\phantom{v} &   93181 &  V4152 Sgr &  9.33\phantom{v} &  1.16\phantom{v} &  0.84\phantom{v} &  0.57\phantom{v} &  0.94\phantom{v} &  1\\
2331\phantom{v} &   43093 &     UZ Pyx &  7.32\phantom{v} &  2.01\phantom{v} &  2.99\phantom{v} &  1.09\phantom{v} &      \phantom{v} &  4 & 4168\phantom{v} &         &            &  9.95\phantom{v} &  1.29\phantom{v} &  1.22\phantom{v} &  0.59\phantom{v} &  1.10\phantom{v} &  4\\
2449\phantom{v} &   45295 &     GM Cnc &  8.65\phantom{v} &  1.57\phantom{v} &  1.50\phantom{v} &  1.00\phantom{v} &  1.93\phantom{v} &  3 & 4179$^{\mathrm{a}}$ &   94049 &            & 10.29\phantom{v} &  1.26\phantom{v} &  0.83\phantom{v} &  0.60\phantom{v} &  1.14\phantom{v} &  3\\
2759\phantom{v} &   50994 &            &  9.53\phantom{v} &  1.30\phantom{v} &  1.07\phantom{v} &  0.59\phantom{v} &  1.09\phantom{v} &  4 & 4194\phantom{v} &   94294 &  V1445 Aql & 11.31\phantom{v} &  2.08\phantom{v} &  2.35: &            1.31\phantom{v} &  2.51\phantom{v} &  3\\
2787\phantom{v} &         &            &  9.48\phantom{v} &  1.29\phantom{v} &  0.96\phantom{v} &  0.60\phantom{v} &  1.11\phantom{v} &  5 & 4196\phantom{v} &         &            & 10.82\phantom{v} &  1.42\phantom{v} &  1.24\phantom{v} &  0.68\phantom{v} &  1.31\phantom{v} &  3\\
2829\phantom{v} &   52271 &            &  7.08\phantom{v} &  1.33\phantom{v} &  1.16\phantom{v} &  0.59\phantom{v} &  1.11\phantom{v} &  4 & 4229\phantom{v} &   94940 &  V1942 Sgr &  7.06\phantom{v} &  2.56\phantom{v} &  4.45: &                \phantom{v} &      \phantom{v} &  1\\
2852\phantom{v} &   52656 &     TZ Car &  8.71v &            2.10\phantom{v} &  2.60\phantom{v} &  1.30\phantom{v} &  2.50\phantom{v} &  4 & 4247\phantom{v} &   95289 &            &  6.96\phantom{v} &  1.07\phantom{v} &  0.58\phantom{v} &  0.57\phantom{v} &  0.97\phantom{v} &  1\\
2925\phantom{v} &   53810 &            &  8.33\phantom{v} &  1.16\phantom{v} &  1.08\phantom{v} &  0.55\phantom{v} &  1.05\phantom{v} &  4 & 4498\phantom{v} &         &            & 11.14\phantom{v} &  1.32\phantom{v} &  1.19\phantom{v} &  0.61\phantom{v} &  1.12\phantom{v} &  3\\
2975\phantom{v} &   54806 &            & 10.16\phantom{v} &  1.44\phantom{v} &  1.14\phantom{v} &  0.85\phantom{v} &  1.64\phantom{v} &  4 & 4524$^{\mathrm{a}}$ &   98117 &            &  9.18\phantom{v} &  1.21\phantom{v} &  0.61\phantom{v} &  0.54\phantom{v} &  1.01\phantom{v} &  3\\
2986\phantom{v} &         &     DI Car & 10.5 v &            1.4 v &            1.30v &            0.64v &            1.2v\phantom{v} &  6 & 4567\phantom{v} &   98223 &            &  9.35\phantom{v} &  2.03\phantom{v} &  2.06\phantom{v} &  0.92\phantom{v} &  1.73\phantom{v} &  4\\
3001\phantom{v} &   55448 &   V905 Cen & 10.51v &            1.80\phantom{v} &  1.87\phantom{v} &  1.15\phantom{v} &  2.20\phantom{v} &  4 & 4595\phantom{v} &   98542 &  V1468 Aql & 10.36\phantom{v} &  2.04\phantom{v} &  2.55\phantom{v} &  1.16\phantom{v} &  2.13\phantom{v} &  3\\
3066\phantom{v} &   56551 &            &  8.76\phantom{v} &  1.06\phantom{v} &  0.51\phantom{v} &  0.51\phantom{v} &  0.92\phantom{v} &  4 & 4598\phantom{v} &   98538 &  V1469 Aql &  8.37\phantom{v} &  2.08\phantom{v} &  2.52\phantom{v} &  0.96\phantom{v} &  1.77\phantom{v} &  3\\
3141\phantom{v} &   58513 &     DD Cru &  8.87\phantom{v} &  2.20\phantom{v} &  2.94: &            1.04\phantom{v} &  2.03\phantom{v} &  4 & 4614\phantom{v} &   98958 &            &  8.05\phantom{v} &  1.07\phantom{v} &  0.97\phantom{v} &  0.51\phantom{v} &  0.98\phantom{v} &  3\\
3199\phantom{v} &         &     TV Cen &  8.02v &            2.74\phantom{v} &  2.89\phantom{v} &  1.42\phantom{v} &  2.57\phantom{v} &  5 & 4873\phantom{v} &  101277 &     BI Cap &  9.67v &            1.42\phantom{v} &  1.09\phantom{v} &  0.95\phantom{v} &  1.85\phantom{v} &  3\\
3227\phantom{v} &   60534 &      S Cen &  7.66v &            1.89\phantom{v} &  2.70: &            1.11\phantom{v} &  2.10\phantom{v} &  5 & 4972\phantom{v} &  102726 &            & 10.30\phantom{v} &  1.29\phantom{v} &  0.92\phantom{v} &  0.63\phantom{v} &  1.14\phantom{v} &  3\\
3286\phantom{v} &   62401 &     RU Vir &  9.97v &            4.63\phantom{v} &  5.10: &            1.99\phantom{v} &  3.42\phantom{v} &  4 & 4978\phantom{v} &  102706 &            &  8.16v &            1.28\phantom{v} &  0.94\phantom{v} &  0.58\phantom{v} &  1.14\phantom{v} &  3\\
3335\phantom{v} &   63955 &            &  8.50\phantom{v} &  1.17\phantom{v} &  1.03\phantom{v} &  0.54\phantom{v} &  1.01\phantom{v} &  4 & 5147\phantom{v} &  104522 &            &  9.82\phantom{v} &  1.56\phantom{v} &  1.49\phantom{v} &  0.97\phantom{v} &  1.86\phantom{v} &  6\\
3405\phantom{v} &   66070 &   V971 Cen &  8.50\phantom{v} &  1.87\phantom{v} &  2.12\phantom{v} &  1.02\phantom{v} &  1.94\phantom{v} &  5 & 5227\phantom{v} &  105212 &            &  9.67\phantom{v} &  1.26\phantom{v} &  0.87\phantom{v} &  0.57\phantom{v} &  1.06\phantom{v} &  5\\
3492\phantom{v} &   70339 &     RS Lup &  9.62v &            2.69\phantom{v} &  4.70: &            1.35\phantom{v} &  2.46\phantom{v} &  5 & 5408\phantom{v} &  107349 &     BU Ind & 10.15v &            1.45\phantom{v} &  1.28\phantom{v} &  0.95\phantom{v} &  1.85\phantom{v} &  4\\
3545\phantom{v} &         &            & 10.95\phantom{v} &  1.40\phantom{v} &  0.80\phantom{v} &  0.77\phantom{v} &  1.44\phantom{v} &  5 & 5420\phantom{v} &  107490 &     RR Ind &  9.34v &            2.84v &            5.29v &            1.31v &            2.36v &            7\\
3558\phantom{v} &         &            & 10.42\phantom{v} &  1.51\phantom{v} &  1.26\phantom{v} &  1.01\phantom{v} &  1.98\phantom{v} &  5 & 5561\phantom{v} &  108953 &     HP Peg &  8.89\phantom{v} &  1.45\phantom{v} &  1.13\phantom{v} &  0.61\phantom{v} &  1.15\phantom{v} &  2\\
3606\phantom{v} &   75694 &     HM Lib &  7.48v &            1.20\phantom{v} &  0.86\phantom{v} &  0.61\phantom{v} &  1.07\phantom{v} &  4 & 5627\phantom{v} &         &            & 10.71\phantom{v} &  1.72\phantom{v} &  1.63\phantom{v} &  0.82\phantom{v} &  1.50\phantom{v} &  3\\
3657\phantom{v} &         &            &  9.84\phantom{v} &  1.59\phantom{v} &  1.32\phantom{v} &  0.69\phantom{v} &  1.28\phantom{v} &  5 & 5761\phantom{v} &  113150 &            & 10.82\phantom{v} &  1.17\phantom{v} &  0.55\phantom{v} &  0.59\phantom{v} &  1.11\phantom{v} &  6\\
3672\phantom{v} &   79484 &            & 10.36\phantom{v} &  1.69\phantom{v} &  1.46\phantom{v} &  0.77\phantom{v} &  1.42\phantom{v} &  5 & 5823\phantom{v} &  114509 &            &  9.26\phantom{v} &  1.22\phantom{v} &  0.81\phantom{v} &  0.60\phantom{v} &  1.11\phantom{v} &  5\\
3707\phantom{v} &   81254 &     LV TrA &  8.30\phantom{v} &  0.95\phantom{v} &  0.67\phantom{v} &  0.45\phantom{v} &  0.72\phantom{v} &  5 & 5937\phantom{v} &  117467 &            &  8.48\phantom{v} &  1.37\phantom{v} &  1.30\phantom{v} &  0.62\phantom{v} &  1.15\phantom{v} &  5\\
3756\phantom{v} &   83387 &      T Ara &  9.03v &            2.78\phantom{v} &  4.90: &            1.40\phantom{v} &  2.55\phantom{v} &  5 & 5980\phantom{v} &     168 &            &  9.55\phantom{v} &  1.12\phantom{v} &  0.48\phantom{v} &  0.51\phantom{v} &  0.96\phantom{v} &  4\\
3765\phantom{v} &         &            &  9.11\phantom{v} &  1.39\phantom{v} &  1.26\phantom{v} &  0.65\phantom{v} &  1.25\phantom{v} &  4 &     \phantom{v} &         &            &      \phantom{v} &      \phantom{v} &      \phantom{v} &      \phantom{v} &      \phantom{v} &   \\
\hline
\end{tabular}
\begin{list}{}{}
\item[$^{\mathrm{a}}$] Close companion: 1871 ($9\arcsec$ separation, bright);
3855 ($15\arcsec$); 3864 ($11\arcsec$); 3966 ($15\arcsec$, bright), 
4179 ($14\arcsec$), 4524 ($13\arcsec$ \&  $18\arcsec$) 
\end{list} 
\end{table*}

\subsection{Photometry of red stars at Siding Spring Observatory}

In March-April 2002 additional $BV\!I_{C}$ photometry for 47 very red Hipparcos 
carbon and M stars was secured at the Siding Spring Observatory, Australia.
The data were obtained using the 24 inch reflector and a single channel
photometer. A cooled unit containing a Hamamatsu GaAs photomultiplier tube 
and a set of filters allow us to match closely the Cousins photometric
system, in the same way as was done at SAAO. Each night a set
of the E-region standards \citep{menzies89} was measured to obtain
the atmospheric extinction coefficients and the transformation
coefficients to the standard system. Mean transformation coefficients
for this run were as follows: $\xi_{V}=0.005$, $\xi_{B-V}=1.010$,
and $\xi_{V-I_{C}}=1.015$ \citep[see][ Eq.~2]{berdnikov01}. Hence
the instrumental system is very close to the standard $BV\!I_{C}$ system,
which greatly alleviates the problem of color-related extrapolation
in the reductions of very red programme stars.
Every 60-90 min two standard stars (red and blue) were used to
define instantaneous zeropoints in the transformation relations. 
Some very bright programme stars were observed with the addition of an  
Oriel 50550 neutral density filter. The $BV\!I_{C}$ photometry 
is presented in Table~\ref{tab:sspring}. 
 
\begin{table}[htb]
\caption[]{\label{tab:sspring} $BV\!I$ photometry at Siding Spring}
\setlength{\tabcolsep}{1mm}
\begin{tabular}{rrcrcc}\hline
HIP & GCVS & JD$-2450000$ & $V$\phantom{.V} & $B\!-\!V$ & $V\!-\!I$ \\ \hline
 23203 &      R Lep & 2353.915 & 11.63 &  4.60 &  3.75\\
 23636 &      T Lep & 2376.883 & 12.18 &  1.75 &  5.92\\
 24055 &      U Dor & 2376.977 &  8.61 &  1.62 &  4.17\\
 &  & 2378.889 &  8.62 &  1.60 &  4.17\\
 25004 &  V1368 Ori & 2376.871 & 10.07 &  3.53 &  3.48\\
 25673 &      S Ori & 2376.874 &  8.71 &  1.65 &  4.66\\
 28041 &      U Ori & 2376.870 & 10.22 &  1.87 &  5.40\\
 &  & 2378.880 &  9.94 &  2.00 &  5.45\\
 29896 &     GK Ori & 2353.919 &  9.96 &  4.22 &  3.52\\
 34413 &      W CMa & 2361.980 &  6.74 &  2.69 &  2.43\\
 35793 &     VY CMa & 2353.922 &  8.19 &  2.28 &  3.28\\
 39967 &     AS Pup & 2376.928 &  9.01 &  1.50 &  4.61\\
 &  & 2378.919 &  9.01 &  1.48 &  4.60\\
 40534 &      R Cnc & 2376.925 & 11.22 &  2.26 &  5.77\\
 &  & 2378.931 & 11.31 &  2.30 &  5.81\\
 41061 &     AC Pup & 2376.908 &  8.99 &  3.23 &  2.78\\
 &  & 2378.933 &  9.04 &  3.31 &  2.80\\
 43905 &      T Cnc & 2353.926 &  8.23 &  4.31 &  3.29\\
 48036 &      R Leo & 2353.929 &  7.28 &  1.71 &  5.02\\
 53085 &      V Hya & 2354.036 &  7.34 &  4.66 &  3.61\\
 53809 &      R Crt & 2354.038 &  8.43 &  2.01 &  4.81\\
 57607 &   V919 Cen & 2354.039 &  6.93 &  1.59 &  4.15\\
 63642 &     RT Vir & 2354.175 &  8.25 &  1.81 &  4.67\\
 64569 &     SW Vir & 2354.178 &  7.09 &  1.72 &  4.53\\
 67419 &      W Hya & 2354.179 &  8.42 &  2.44 &  5.64\\
 69754 &      R Cen & 2354.178 &  7.48 &  1.94 &  4.22\\
 70969 &      Y Cen & 2354.181 &  8.12 &  1.60 &  4.50\\
 75393 &     RS Lib & 2354.182 & 10.79 &  1.96 &  5.41\\
 80365 &     RT Nor & 2354.183 & 10.08 &  1.01 &  0.94\\
 80488 &      U Her & 2379.168 &  8.68 &  1.60 &  4.84\\
 80550 &      V Oph & 2357.168 &  9.21 &  4.13 &  3.19\\
 82392 &      V TrA & 2364.266 &  8.16 &  2.23 &  2.24\\
 84876 &  V1079 Sco & 2354.185 &  9.40 &  3.31 &  3.34\\
 85617 &     TW Oph & 2357.168 &  7.86 &  4.24 &  3.32\\
 85750 &            & 2357.174 &  9.36 &  1.93 &  1.65\\
 86873 &     SZ Sgr & 2357.170 &  8.78 &  2.36 &  2.73\\
 87063 &     SX Sco & 2357.172 &  7.65 &  2.86 &  2.65\\
 88341 &  V4378 Sgr & 2379.172 & 10.37 &  2.97 &  3.24\\
 88838 &     VX Sgr & 2379.174 &  9.20 &  2.82 &  4.34\\
 89739 &     RS Tel & 2357.254 & 10.01 &  0.85 &  0.77\\
 90694 &            & 2357.252 &  9.93 &  1.37 &  1.61\\
 93605 &     SU Sgr & 2357.258 &  8.33 &  1.73 &  4.39\\
 93666 &      V Aql & 2357.260 &  6.78 &  3.98 &  3.07\\
 98031 &      S Pav & 2379.201 &  7.82 &  1.64 &  4.63\\
 99082 &  V1943 Sgr & 2379.195 &  7.67 &  1.77 &  4.58\\
 99512 &      X Pav & 2357.256 &  8.97 &  1.91 &  4.92\\
100935 &      T Mic & 2357.265 &  7.68 &  1.78 &  4.76\\
\hline
\end{tabular}
\end{table}

\subsection{$V\!RI$ photometry of red variables with APT}

Since 1996 the University of Vienna has been obtaining $U\!BV\!(RI)_{C}$
photometry in Arizona using two 0.75m automatic photoelectric 
telescopes\footnote{operated by the University of Vienna and 
the Astrophysikalisches Institut, Potsdam} (APT)
located on the grounds of Fairborn Observatory. 
The photometer of the APT dubbed Amadeus \citep{strassmeier97}, has 
an EMI-9828 S-20/B multi-alkali cathode photomultiplier,
which is sensitive up to $\sim900$ nm. This photomultiplier in combination
with filters close to those suggested by \citet{bessell76} 
reproduces a $V\!(RI)_{C}$ system close to the one used by \citet{walker79}.   
In 1997 a monitoring programme of nearly 60 late spectral type semiregular and
irregular variables was initiated.  Typical light curves resulting from
this programme can be found in \citet{lebzelter99} and
\citet{kerschbaum01}. A complete sample of light curves will be published 
elsewhere (Lebzelter et al., in preparation). In Table~\ref{tab:apt}
we present median $V$, $V\!-\!I_{C}$, and an intercept $a_0$ and slope $a_1$ 
from the fit $V\!-\!I$ vs. $V$ for 45 selected Hipparcos variables used in the
following calibration (Sect.~\ref{calib}). The total number of observations
$n$ is indicated in the last column.

\begin{table}[htb]
\caption[]{\label{tab:apt} APT photometry of selected red variables}
\setlength{\tabcolsep}{1mm}
\begin{tabular}{rrccrrr}\hline
HIP & GCVS & $V$  & $V\!-\!I_{C}$ & $a_0$\phantom{a} & $a_1$\phantom{a} & 
$n$ \\ \hline
  4008 & VY Cas     &    9.49 &    4.14 &    0.66 &  0.366 & 217\\
  5914 & Z Psc      &    6.85 &    2.54 &   $-0.18$ &  0.396 &  49\\
  6191 & AA Cas     &    8.24 &    3.47 &    0.00 &  0.422 & 206\\
 10472 & V Ari      &    8.52 &    2.07 &   $-1.15$ &  0.379 &  30\\
 17821 & BR Eri     &    7.15 &    3.16 &   $-0.15$ &  0.465 & 270\\
 21046 & RV Cam     &    8.16 &    3.81 &    0.38 &  0.420 & 326\\
 22667 & $o^{1}$ Ori  &    4.84 &    2.50 &  $-0.10$ &  0.536 &  83\\
 32083 & VW Gem     &    8.32 &    2.41 &   $-0.86$ &  0.391 &  36\\
 33369 & BG Mon     &    9.66 &    2.46 &   $-1.40$ &  0.400 &  35\\
 41061 & AC Pup     &    9.05 &    2.83 &   $-1.42$ &  0.474 & 360\\
 41201 & FK Hya     &    7.29 &    3.48 &    0.22 &  0.446 & 388\\
 43063 & EY Hya     &    9.60 &    4.49 &    1.01 &  0.366 &  85\\
 44601 & TT UMa     &    9.02 &    3.68 &   $-0.17$ &  0.427 & 425\\
 44862 & CW Cnc     &    8.70 &    4.03 &    0.90 &  0.360 &  67\\
 56976 & AK Leo     &    8.54 &    2.87 &   $-1.37$ &  0.497 &  68\\
 57504 & AZ UMa     &    8.50 &    3.97 &    0.57 &  0.400 & 440\\
 59108 & RW Vir     &    7.33 &    3.63 &    0.66 &  0.405 & 377\\
 61022 & BK Vir     &    7.81 &    4.24 &    2.13 &  0.268 &  98\\
 61839 & Y UMa      &    8.39 &    4.40 &    1.92 &  0.295 & 411\\
 66562 & V UMi      &    7.91 &    2.92 &   $-0.95$ &  0.488 &  78\\
 69449 & EV Vir     &    6.91 &    2.62 &   $-1.05$ &  0.533 & 223\\
 70236 & CI Boo     &    6.48 &    2.93 &   $-0.63$ &  0.549 & 182\\
 70401 & RX Boo     &    7.43 &    4.33 &    2.97 &  0.184 & 105\\
 71644 & RV Boo     &    8.24 &    4.06 &    1.30 &  0.333 & 190\\
 73213 & FY Lib     &    7.24 &    3.65 &    0.30 &  0.460 & 225\\
 74982 & FZ Lib     &    7.10 &    3.04 &   $-1.00$ &  0.570 & 367\\
 78574 & X Her      &    6.28 &    3.92 &    1.60 &  0.371 & 346\\
 80259 & RY CrB     &    9.63 &    4.02 &    0.24 &  0.393 & 255\\
 80704 & g Her      &    4.86 &    3.47 &    1.23 &  0.461 & 291\\
 81188 & TX Dra     &    7.26 &    2.96 &   $-0.58$ &  0.488 & 153\\
 81747 & AX Sco     &    8.73 &    4.00 &   $-0.60$ &  0.527 & 120\\
 82249 & AH Dra     &    7.54 &    3.52 &    0.00 &  0.465 & 301\\
 84027 & CX Her     &    9.86 &    4.04 &    1.85 &  0.225 &  33\\
 84329 & UW Her     &    7.97 &    3.42 &   $-0.29$ &  0.464 & 298\\
 84346 & V438 Oph   &    9.12 &    4.26 &    2.41 &  0.199 & 164\\
 93989 & V398 Lyr   &    7.39 &    3.30 &   $-0.32$ &  0.490 & 265\\
 95173 & T Sge      &    9.29 &    4.66 &    2.45 &  0.236 & 276\\
 96919 & V1351 Cyg  &    6.56 &    3.06 &    0.00 &  0.466 & 226\\
102440 & U Del      &    6.77 &    3.61 &    0.88 &  0.402 & 299\\
103933 & DY Vul     &    7.09 &    3.58 &    0.55 &  0.425 & 207\\
107516 & EP Aqr     &    6.63 &    4.01 &    2.16 &  0.279 & 183\\
109070 & SV Peg     &    8.67 &    4.47 &    0.18 &  0.490 &  69\\
110099 & UW Peg     &    8.89 &    3.39 &   $-0.82$ &  0.473 & 207\\
112155 & BD Peg     &    8.66 &    3.82 &    0.56 &  0.376 & 159\\
113173 & GO Peg     &    7.37 &    2.66 &   $-0.76$ &  0.464 & 168\\
\hline
\end{tabular}
\end{table}

\subsection{Published sources of $V\!I$ photometry}

Only two large surveys of relatively bright
red stars are available in the $V\!I_{C}$
system -- a survey of the Southern carbon stars \citep{walker79}
and the recent photometry of nearly 550 Hipparcos M stars \citep{koen02}.
Additional literature on the $V\!I_{C}$ photometry of Hipparcos 
red stars is not rich,
therefore we included some other sources containing Johnson $V\!I_{J}$
photometry. We used normal color indices for M0 to M8 spectral type stars
\citep[][ Table~4]{celis86} to obtain the following relationship 
between the Johnson $V\!-\!I_{J}$ and Cousins $V\!-\!I_{C}$:
\begin{equation}
V\!-\!I_{C}=-0.359+0.894(V\!-\!I)_{J}-0.0087(V\!-\!I)_{J}^{2},
\end{equation}
defined for the giants of M spectral type. This is valid for zirconium 
(S-type) stars, and probably usable for carbon stars as well,
throughout the $V\!-\!I_{J}$ range from 1.9 to 8.7 mag. 
Note that this relationship yields a bluer color index, by $\sim\!0.1$,
than a similar relationship from Hipparcos Catalogue \citep[][ vol.1] {esa97}.
A list of all sources used in this paper to calibrate $V\!-\!I$ photometry
is given in Table~\ref{tab:all}. It contains the reference, the number of stars
$n$, spectral type, photometric system, and remarks. This list is not
complete since we deliberately left out a few sources for the further
independent comparisons.

\begin{table}[htb]
\caption[]{\label{tab:all} Selected sources of $V\!I$ photometry}
\setlength{\tabcolsep}{1mm}
\begin{tabular}{lrcll}\hline
Source & $n$ & Type & System & Remarks \\ \hline
\citet{bagnulo98}   &   1 & C   & Cousins &  \\
\citet{barnes73}    &  11 & M   & Johnson & narrow-band $I$\\
\citet{celis82}     &  24 & M   & Kron(?) & $\sim$Johnson $I$\\
\citet{celis86}     &  20 & M   & Cousins &  \\
\citet{eggen72}     &  30 & C   & Eggen   & $\sim$Cousins $I$\\
\citet{delaverny97} &   2 & C   & Cousins & \\
\citet{kizla81}     &  36 & C,M & Johnson & \\
\citet{koen02}      &  80 & M   & Cousins &  only $V<8.4$\\
\citet{lee70}       &  43 & M   & Johnson & \\
\citet{mendoza65}   &  33 & C   & Johnson & \\
\citet{olson75}     &  11 & C   & Johnson & \\
\citet{percy01}     &  16 & C,M & Johnson & \\
\citet{walker79}    & 119 & C   & Cousins & \\
Table~\ref{tab:evans}  &  61 & C   & Cousins & this study \\
Table~\ref{tab:sspring}  &  42 & C,M & Cousins & this study \\
Table~\ref{tab:apt}   &  45 & C,M & Cousins & this study \\
\hline
\end{tabular}
\end{table}

\subsection{Radial velocities} 

Although radial velocities have no direct bearing on the photometry,
they could be used to identify spectroscopic binaries and hence
shed light on possible discrepancies in the photometry caused by 
duplicity. We selected 19 Hipparcos carbon stars, mostly R type. 
The radial velocity measurements were made with a Coravel-type
spectrometer using the Steward Observatory 1.6m Kuiper Telescope 
at Mt. Bigelow, Arizona in February, 2002. Additional measurements
were also obtained with the Moletai Observatory 1.65m telescope in
Lithuania and the 1.5m telescope of the Turkish National Observatory
near Antalya. A detailed description of the spectrometer is given
in \citet{upgren02}. On average, the estimated precision of a single 
measurement is 0.7 km s$^{-1}$. A total of 61 measurements of radial 
velocity are given in Table~\ref{tab:rv}, where columns 1-6 are Hipparcos
number, carbon star number from \citet{stephenson89},
GCVS variable star name \citep{kholopov85},  Julian date, heliocentric 
radial velocity and
its estimated standard error, both in km s$^{-1}$. More details on 
the observing and reduction procedure can be found in \citet{upgren02}. 
By examining the ratio of external and internal error in accordance
with \citet{jasniewicz88}, it is evident that two stars in 
Table~\ref{tab:rv}, HIP 53522 and 53832, are new SB1 spectroscopic binaries,
although the time span is too short for the orbit determination. 
Both stars are suspected CH-like carbon stars \citep{hartwick85}, 
which adds more weight to the paradigm that most CH stars are binaries.

\begin{table*}[htb]
\caption[]{\label{tab:rv} Radial velocities of R and other selected 
carbon stars}
\setlength{\tabcolsep}{1mm}
\begin{tabular}{rcrcrc|rcrcrc}\hline
HIP & CGCS & GCVS\phantom{SS} & JD-2450000 & RV$_{\rm hel}$ & $\sigma_{\rm RV}$ & 
HIP & CGCS & GCVS\phantom{SS} & JD-2450000 & RV$_{\rm hel}$ & $\sigma_{\rm RV}$
\\ \hline
 26927 & 1035 & $\cdots$\phantom{SSS} & 2327.617 & 42.5 & 0.6 & 53832 & 2919 & $\cdots$\phantom{SSS} & 2327.946 &  5.2  &   0.7  \\ 
       &      & & 2332.630 & 42.2 & 0.6 &       & &  & 2332.843  &    3.4  &   0.6  \\ 
 29896 & 1222 &     GK Ori & 2330.729  &  54.6  &   1.5&       & &  & 2363.492  &   $-2.8$ &  0.7  \\
 $\cdots$  &1226&V1393 Ori$^{\mathrm{a}}$ &2332.641&34.2 &   0.6&       & &  & 2368.389  &   $-3.1$ &  0.7  \\
 29961 & 1230 &  V1394 Ori & 2327.658  &   70.8  &   0.7&       & &  & 2382.344  &   $-5.5$ &  0.7 \\
 31829 & 1337  & NY Gem & 2327.732  & $-123.0$  &   0.8&       & &  & 2386.347  &   $-6.4$ &  0.7  \\
 32187 & 1373 & V738 Mon & 2327.706  &   60.3  &   0.7&       & &  & 2399.392  &   $-8.9$ &  0.7  \\
       & &            & 2332.650  &   61.2  &   0.7&       & &  & 2403.329  &   $-8.4$ &  0.7  \\
 33369 & 1474 &  BG Mon & 2327.752  &   71.4  &   0.7&       & &  & 2419.261  &   $-11.5$ & 0.7  \\
       & &            & 2333.745  &   71.6  &   0.7&       & &  & 2423.270  &   $-12.1$ & 0.7  \\
       & &            & 2350.255  &   71.4  &   0.7& 58786 & 3156& $\cdots$\phantom{SSS}  & 2349.500 & $-21.3$ & 0.7  \\
 34413 & 1565 &      W CMa & 2330.686  &   18.9  &   0.6&       & &  & 2368.400  &   $-21.4$ & 0.7  \\
       & &            & 2333.737  &   19.6  &   0.6&       & &  & 2386.299  &   $-21.2$ & 0.7  \\
 35681 & 1622 &     RU Cam & 2350.266  &  $-24.4$  &   0.6& 62944 & $\cdots$  & $\cdots$\phantom{SSS}   & 2327.992  &    8.5  &   0.6  \\
       & &            & 2356.253  &  $-26.3$  &   0.7&       & &  & 2332.853  &    6.5  &   0.6  \\
       & &            & 2375.335  &  $-24.9$  &   0.7&       & &  & 2363.504  &    6.7  &   0.6  \\
 38242 & 1891 & $\cdots$\phantom{SSS}   & 2327.760  &   13.7  &   0.7&       & &  & 2368.416  &    6.2  &   0.6  \\
       & &  & 2332.664  &   15.7  &   0.7&       & &  & 2382.382  &    7.3  &   0.6  \\
 39118 & 1981 & $\cdots$\phantom{SSS}   & 2327.772  &   95.5  &   0.7&  63955 & 3335 & $\cdots$\phantom{SSS}  & 2327.957 &$-9.2$ &  0.7 \\
       & &  & 2332.670  &   96.2  &   0.7&       & &  & 2332.919  &  $-10.1$ &  0.6  \\
 44812 & 2428 & $\cdots$\phantom{SSS}   & 2327.917  &   20.2  &   0.7& 69089 & $\cdots$  & $\cdots$\phantom{SSS}   & 2330.980  &  $-20.3$ &  0.6  \\
       & &  & 2350.314  &   20.1  &   0.8&       & &  & 2332.906  &  $-20.3$ &  0.7  \\
       & &  & 2375.300  &   20.7  &   0.7&       & &  & 2359.572  &  $-21.4$ &  0.6  \\
 50412 & 2715 & $\cdots$\phantom{SSS}   & 2349.486  &  $-84.8$  &   0.7&       & &  & 2382.449  &  $-20.2$ &  0.6  \\
       & &  & 2386.284  &  $-84.9$  &   0.7&       & &  & 2399.479  &  $-20.9$ &  0.6  \\
 53354 & 2892 &  $\cdots$\phantom{SSS}  & 2330.801  &    4.7  &   0.7& &&&&&\\
       & &  & 2332.827  &    5.8  &   0.8& &&&&&\\
 53522 & 2900 & $\cdots$\phantom{SSS}  & 2327.938  &   28.0  &   0.6& &&&&&\\
       & &  & 2369.333  &   31.5  &   0.7& &&&&& \\
       & &  & 2375.326  &   33.3  &   0.7& &&&&& \\
       & &  & 2382.336  &   34.9  &   0.7& &&&&& \\
       & &  & 2386.337  &   34.3  &   0.7&  &&&&&\\
       & &  & 2399.377  &   36.3  &   0.7&  &&&&&\\
       & &  & 2403.322  &   37.5  &   0.7&  &&&&&\\
       & &  & 2419.255  &   37.5  &   0.7&  &&&&&\\
       & &  & 2423.255  &   38.5  &   0.7&  &&&&&\\
\hline
\end{tabular}
\begin{list}{}{}
\item[$^{\mathrm{a}}$] Not HIP 29899. See Table~\ref{tab:off}.
\end{list} 
\end{table*}

\section{Deriving $V\!-\!I$ from Hipparcos~$Hp$}\label{calib}

The central idea of this study is to derive new sets of $V\!-\!I$ color
indices for
red stars bypassing all various methods used in the original derivation
of $V\!-\!I$ \citep{esa97}. We abandon the calibration methods based upon the
ground-based $B\!-\!V$ or Tycho $B_{T}\!-\!V_{T}$ for two reasons. First, the
$B\!-\!V$ color index, at least for carbon stars, is a poor representative 
of effective temperature due to the severe blanketing effect by molecular 
bands \citep{alksne91} in the $BV$ bandpasses.
Second, many Hipparcos red stars have such a large
$B\!-\!V$ color index that their measurements are uncertain or, in the case of
Tycho magnitudes, missing due to extremely low fluxes in the $B_{T}$
bandpass. In this sense the potential of Tycho $B_{T}V_{T}$ photometry
for red stars is limited. However, there is a color index, $Hp-V_{T}$,
which to our knowledge, has been used neither in the Hipparcos
reductions nor the following studies. 
\begin{figure}[htb]
\resizebox{\hsize}{!}{\includegraphics{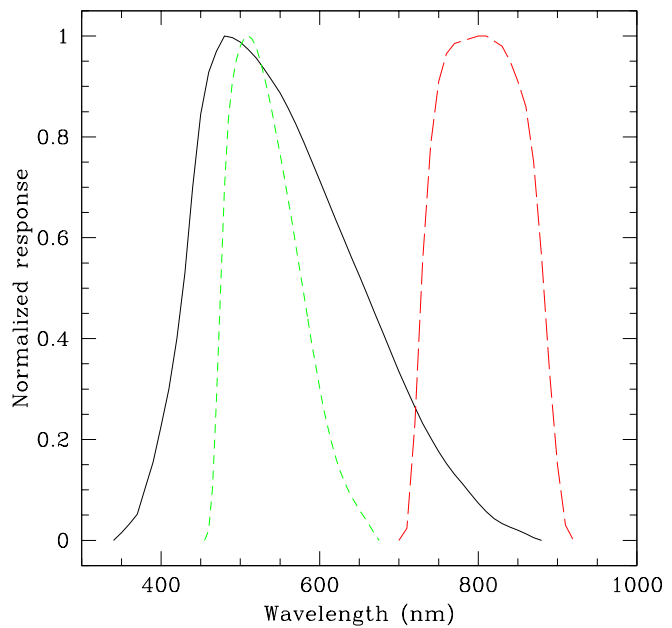}}
\caption[]{\label{fig:band} Normalized response curves for the Hipparcos
$Hp$ (solid line), Tycho $V_{T}$ (short-dashed line), and Cousins $I$
(long-dashed line) bandpasses. The corresponding curves are taken from
\citet{bessell90,bessell00}.
}
\end{figure}
The normalized $Hp$ and $V_{T}$
response curves provided by \citet{bessell00} indicate only a 21 nm 
difference in the mean wavelength (see Fig.~\ref{fig:band}).
This wavelength is calculated assuming a flat spectral energy distribution
(SED) which is definitely not the case for late-type stars. If we account
for the observed spectral energy distribution, e. g., from \citet{gunn83},
then for an M7III spectral-type star (HIP 64569) 
the difference in the effective wavelengths of the two filters reaches 150 nm. 
The SEDs for the two carbon stars HIP 99 and 95777 
yield an 84 and 94 nm difference in the effective wavelength, respectively.
It is the extended red response of the S20 photocathode of Hipparcos main
detector -- Image Dissector Tube, which makes the $Hp-V_{T}$ index
fairly sensitive in the K-M spectral range 
\citep[see][ vol.~1,~Fig.~1.3.4]{esa97}.  We employ this property
to calibrate $V\!-\!I$ for late-type stars using $Hp-V_{T}$. 
 
\subsection{Tycho photometry}

\begin{figure}[htb]
\resizebox{\hsize}{!}{\includegraphics{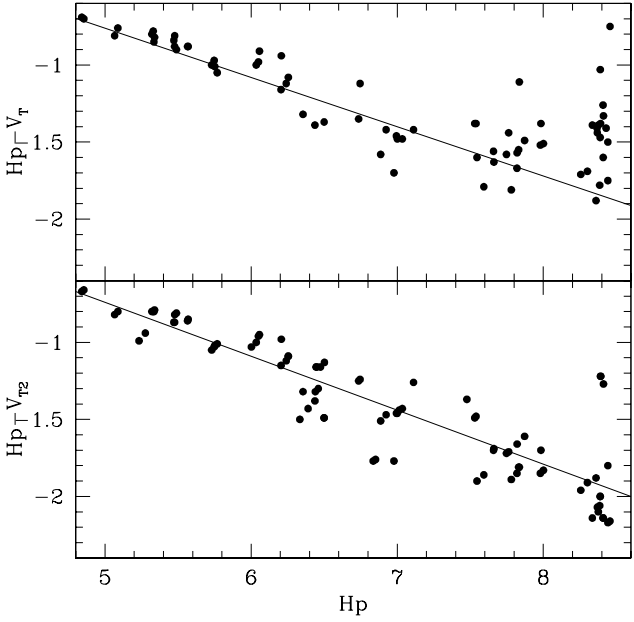}}
\caption[]{\label{fig:bias} Bias in the $Hp-V_{T}$ at $Hp>8$ originating from
original Tycho $V_{T}$ magnitudes for a bright Mira T Cep = HIP 104451
(top panel). If the $V_{T2}$ epoch photometry is used, the
bias disappears (bottom panel).  A straight line is fitted to the data
in the bottom panel and then just overplotted in the top panel.
}
\end{figure}

First trials using the published Tycho $V_{T}$ photometry indicated 
two problems. First, a large fraction of red stars lack Tycho
photometry. Second, the $V_{T}$ photometry shows a progressively 
increasing bias at faint magnitudes ($V_{T}>9$). This effect is
illustrated by Fig.~\ref{fig:bias} where $Hp-V_{T}$ values are 
abnormally small at $Hp>8$, equivalent to the `brightening' of 
$V_{T}$ at these $Hp$ magnitudes. 
It is suspected that the de-censoring
technique \citep{halbwachs97} has failed to completely correct the
faint-magnitude bias. Therefore, it was decided to make use of the 
Identified Counts Data Base, ICDB \citep{fabricius00} -- a by-product 
of the Tycho-2 data re-processing \citep{hog00}.

All transits of about 2.5 million stars included in the
Tycho-2 Catalogue are represented in the ICDB by sequences of
13 time-ordered photon
counts, separately for the inclined and vertical slits, and the
$B_T$ and $V_{T}$ bandpasses. Combined with some instrument calibration
files, this data base is sufficient to reproduce a complete astrometric
solution for any Tycho-2 star, including its possible binarity status,
photometric variability, etc. In this paper, we exploit the
possibility to extract epoch photometry for selected stars by estimating 
the signal at the pre-computed, mission-averaged astrometric position.

The working version of Tycho-2 epoch photometry was derived some time ago
for a search of a particular kind of variable stars, although 
it has not been implemented in the construction of the Tycho-2 Catalogue. 
It should be noted that, even though based on the same observational data, 
the Tycho-2 epoch photometry used here differs significantly from
the published Tycho epoch photometry \citep{esa97}. Nevertheless,
the global calibrations of our current epoch photometry are consistent
with the Tycho mission-average calibrations. On the star-by-star level,
the Tycho-2 processing (both astrometric and photometric) is based on a 
single so-called Maximum Cross-Correlation estimator, while the original 
Tycho epoch photometry is the result of a series of successive linear and 
non-linear filterings \citep[ vol.~4]{halbwachs97,esa97}. The
main difference in the reduction procedure is that for a given star in 
Tycho-2, the determination of astrometric parameters was done over 
all collected transits at once; whereas in Tycho, a complete cycle of 
astrometric and photometric reductions was performed for each transit.
 
The latter method proved to be unreliable at a low signal-to-noise ratio, 
as the noise may mimic a signal from the star and 
produce a spurious astrometric detection and a subsequent false
photometric estimate at the derived location. Such false detections
tend to be abnormally bright, which then produce a bias in the 
faint magnitudes and hence necessitate the de-censoring analysis 
\citep{halbwachs97} as the lesser of two evils.

The Tycho-2 epoch photometry is largely free of this de-censoring bias,
since all photometric estimations are made at the correct location of a star
image (within the astrometric precision), and all observations are retained.
Still, Tycho-2 epoch photometry can only find restricted applications
due to a possibly high background and contamination from other stars
which could be present in the $40\arcmin$-long slits of the star mapper.

We will denote the re-processed Tycho photometry as $V_{T2}$ to
distinguish it from the original Tycho $V_{T}$ epoch photometry.

\subsection{Relationship $Hp\!-\!V_{T2}$ vs. $Hp$}\label{slopes}

Due to the differences in spectral features, we kept the processing
of carbon and oxygen- and zirconium-rich (M, S) stars separately.
There are 321 carbon stars and 4464 stars of M and S spectral type, which
have a pair of $Hp$ and $V_{T2}$ values. These
stars were selected according to the listed spectral type in the
Hipparcos Catalogue (field H76) but not fainter than $Hp=11$. In the case
of a missing spectral type, we included the stars having Hipparcos
$V\!-\!I>1.5$. Finally, the stars of K spectral-type were also considered if 
their $V\!-\!I>2$. Note that for the Hipparcos photometry we used the 
so-called $Hp_{\rm dc}$ magnitude 
estimate derived from the unmodulated part of a signal intensity 
\citep{esa97}, since the mean photometric parameters have been obtained 
from  $Hp_{\rm dc}$. In addition, the ground-based photoelectric photometry 
is always integrated over some aperture (usually with
$\diameter=15-30\arcsec$) centered onto the target and hence, the flux 
from any object within this aperture is going to be included. 
However, in Tycho-2 photometry, if the star was found 
to be a binary (minimum separation $\sim0\farcs4$), only the brightest 
component has been retained and subsequently used for this study.
Because of that, the color index $Hp-V_{T2}$ of resolved binaries could be 
biased to some degree and thus, should be considered with caution.

For each
star, the color index $Hp-V_{T2}$ was visually examined as a function of $Hp$
ignoring the listed status flags. A pair of $Hp,V_{T2}$ photometry 
was deleted if it deviated
 from the mean trend by more than $3\sigma$. 
\begin{figure}[htb]
\resizebox{\hsize}{!}{\includegraphics{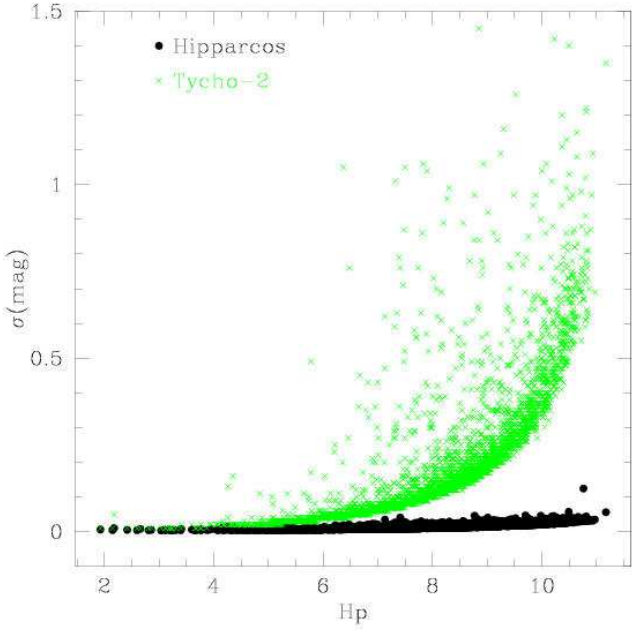}}
\caption[]{\label{fig:err} The distribution of mean standard errors for
red stars from $Hp$ photometry (solid dots) and Tycho-2 $V_{T2}$ data 
(crosses) as a function of median $Hp$ magnitude.  The large scatter in 
the distribution of $V_{T2}$ errors is due to the variability -- 
observations of Miras generate the largest scatter. The lower envelope of
the same error distribution reflects the contribution by photon noise.
}
\end{figure}
As seen in Fig.~\ref{fig:err} 
the precision of $Hp-V_{T2}$ is driven by the precision of the
$V_{T2}$ photometry. A rapidly deteriorating error budget at $Hp>9$
actually poses a problem of reliability of calculated slopes in the 
$Hp-V_{T2}$ vs. $Hp$ plot. We opted for an interactive and iterative
linear fit to find a slope, i.e., gradient  
$\nabla_{HpV_{T}}=\Delta(Hp-V_{T2})/\Delta{Hp}$ and an intercept.
It was decided to keep all datapoints unless any were clearly 
deviant or there was a peculiar trend usually due to very faint 
or corrupted $V_{T2}$ epoch photometry. 
It should be noted that we were not able to find a perceptible
difference in the color of variable stars observed at the same 
magnitude on the ascending or descending part of a lightcurve. 
In the case of a constant star or large uncertainties in the 
$V_{T2}$ photometry, only the mean $Hp-V_{T2}$ has been calculated. 
We note that $Hp$ can be predicted for any 
$V_{T2}$ via 
\begin{equation}
Hp={b_0+V_{T2} \over 1-b_1}, \label{eq:hp}
\end{equation}
where $b_0$ is the intercept and $b_1$ is the slope from a linear fit.
This simple relationship is crucial in bridging the ground-based 
$V\!I$ photometry and Hipparcos $Hp$ photometry (see Sect.~\ref{sect:cal}). 

\begin{figure}[htb]
\resizebox{\hsize}{!}{\includegraphics{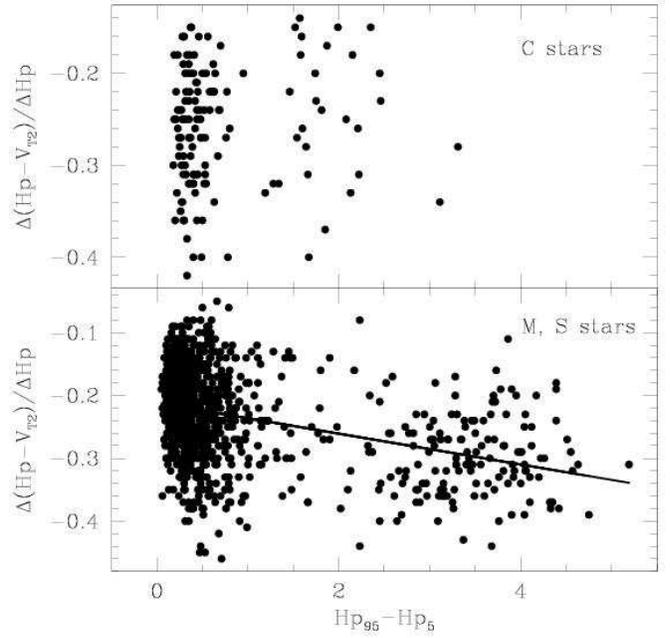}}
\caption[]{\label{fig:slopes} Calculated gradients 
$\Delta (Hp-V_{T2})/\Delta Hp$ as a function of observed $Hp$ amplitude
within the 5-to-95 percentile range, $Hp_{95}-Hp_{5}$, for 136 carbon stars
(top panel) and 906 M and S stars (bottom panel). For M and S stars, the 
gradient is correlated with the amplitude of the brightness variation as 
indicated by a linear fit.
}
\end{figure}
The calculated color gradients $\nabla_{HpV_{T}}$ 
vs. the observed amplitude in $Hp$ within the 5-to-95 percentile range,
 $Hp_{95}-Hp_{5}$, are shown 
in Fig.~\ref{fig:slopes}, separately for 136 carbon and 906 M and S stars. 
For both groups of stars, the color gradient ranges between $-0.1$ and $-0.45$.
For carbon stars, the mean gradient is $\langle\nabla_{HpV_{T}}\rangle=\!-0.24$,
whereas it is $-0.26$ for the M and S stars. This indicates that on average
the gradient $\nabla_{HpV_{T}}$ is only marginally sensitive to the
C/O ratio in the atmospheres of red stars. On the other hand, for M and S
stars, the gradient is definitely correlated with the amplitude
of a brightness variation in $Hp$ -- the color gradient increases at the rate
$-0.025$ per mag of amplitude. Similarly, the gradient is correlated with 
the median $V\!-\!I$ for M and S stars: this merely reflects another correlation
between the amplitude of brightness variation and median $V\!-\!I$.

\subsection{$V\!-\!I$ calibration curves}\label{sect:cal}

We have not been able to find any ground-based $V\!-\!I_{C}$ data for the red
stars concurrent with the Hipparcos lifetime. To relate the ground-based 
$V\!-\!I$ observations to Hipparcos/Tycho photometry we postulate that a 
star's luminosity-color relation (encapsulated by parameters $b_0$ and 
$b_1$ in Eq.~\ref{eq:hp}) is constant over several decades and
adopt the $V_{T2}$ magnitude as a proxy to tie ground-based observations
into the Hipparcos $HpV_{T2}$ system. In practice, it
involves two important steps. First, the ground based $V$ magnitude should
be transformed into the system of Tycho $V_{T}$. This is not trivial for
red stars, therefore we provide step-by-step instructions explaining how 
to do that for carbon and M, S stars. Second, the derived 
$V_{T2}$ magnitude now allows us to find the corresponding $Hp$ value using 
Eq.~\ref{eq:hp} and thus, the color $Hp-V_{T2}$. Only then, it is possible to 
relate a ground-based measurement of $V\!-\!I$ to the corresponding 
$Hp-V_{T2}$ value and be reasonably certain that both measurements are
on the same phase of a light curve in the case of variable stars. 
As demonstrated by 
\citet{kerschbaum01}, there is no phase shift between the variability 
in the $V$ and $I_C$ bandpasses for asymptotic giant branch stars, a dozen
of which can also be found in Table~\ref{tab:apt}. A small and consistent rms 
scatter of the residuals in the linear fits given in Table~\ref{tab:apt} for 
additional M stars and a few carbon stars, is another reassuring sign of 
the lack of a phase shift -- a crucial assumption in the calibration procedure. 

\subsubsection{Carbon stars}

Many carbon stars are too faint in the $B_T$ bandpass, hence their $B_{T}-V_{T}$
color index is either unreliable or is not available at all. Therefore, we 
first derived a relationship between the ground-based $(V\!-\!I)_{C}$ 
and $(B-V)_{J}$ using the \citet{walker79} data: 
\begin{equation}
(B-V)_{J}=1.59-0.942(V\!-\!I)_{C}+0.5561(V\!-\!I)_{C}^{2}.
\end{equation}
Then, the $B_{T}-V_{T}$ can be easily estimated
using Eq. 1.3.31 in ESA (1997), vol. 1:
\begin{equation}
(B_{T}-V_{T})=1.37(B-V)_{J}-0.26.
\end{equation}
Finally, knowing the ground-based $V$-magnitude and employing Eq. 1.3.34
in ESA (1997), vol. 1, we derive 
\begin{equation}
V_{T2}=V_{J}-0.007+0.024(B_{T}-V_{T})+0.023(B_{T}-V_{T})^{2},
\end{equation}
which  in combination with Eq.~\ref{eq:hp} yields the corresponding $Hp-V_{T2}$.

\subsubsection{M and S stars}

Owing to some, albeit 
 weak, dependence of TiO absorption upon the surface gravity, the stars of
spectral type M can be divided into giants and dwarfs (main sequence stars).
All stars in our sample with Hipparcos parallaxes smaller than 10 mas are
considered to be giants. For M giants, $V_{T2}$ follows directly from 
Eq. 1.3.36 (see ESA 1997, vol. 1):
\begin{equation}
V_{T2}=V_{J}+0.20+0.03(V\!-\!I-2.15)+0.011(V\!-\!I-2.15)^{2}. \label{eq:giants}
\end{equation}
To calculate a similar relationship for M dwarfs, we used the data from
\citet{koen02}:
\begin{equation}
V_{T2}=V_{J}+0.20+0.042(V\!-\!I-2.15). \label{eq:dwarfs}
\end{equation}
As expected, Eqs.~\ref{eq:giants}\&\ref{eq:dwarfs} are very similar so
that, considering the uncertainties involved, our $V\!-\!I$ photometry is not 
sensitive to the surface gravity.  
Eq.~\ref{eq:giants} or \ref{eq:dwarfs} in combination with 
Eq.~\ref{eq:hp} then yields $Hp-V_{T2}$.

\subsubsection{Calibration curves}

From the sources listed in Table~\ref{tab:all}, we have chosen 274 
measurements of $V\!-\!I$ for carbon stars and 252 for M and S stars. 
Quite often there
is more than one $V\!-\!I$ measurement for a given star. In the case of
multi-epoch ground-based $V\!-\!I$ data, we first obtained a linear
fit to $V\!-\!I$ as a function of $V$ (e.g., Table~\ref{tab:apt}). 
The coefficients of that fit were used to estimate the 
$V\!-\!I$ index of variable stars at maximum brightness. 
The corresponding $Hp\!-\!V_{T2}$ color index at maximum brightness has the
advantage of being relatively insensitive to the uncertainties
affecting the $Hp\!-\!V_{T2}$ vs. $Hp$ relation at its faint end
(see Figs.~\ref{fig:bias},\ref{fig:err}).   
This is especially important at the
blue end of the relationship between $V\!-\!I$ and $Hp\!-\!V_{T2}$ 
(corresponding to the maximum brightness in the case of
variable stars) requires more care due to its steepness.

\begin{figure}[htb]
\resizebox{\hsize}{!}{\includegraphics{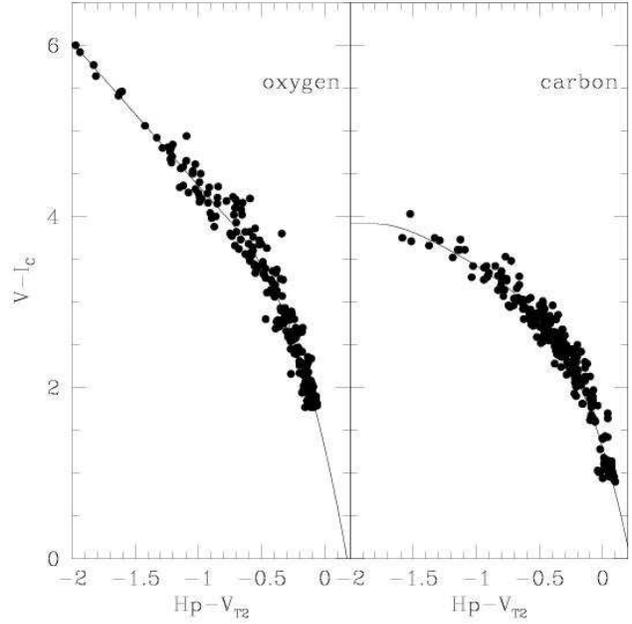}}
\caption[]{\label{fig:coltran} 
Color-color transformation for M and S stars (left panel) and carbon
stars (right panel). The red end of this transformation 
($Hp\!-\!V_{T2}\!<\!-1.5$)
for carbon stars is uncertain due to the lack of intrinsically very red 
Hipparcos calibrating carbon stars.
}
\end{figure}

The calibration curves for oxygen (actually
M and S) stars and carbon stars are presented in Fig.~\ref{fig:coltran}.  

Since many calibrating stars are fainter than
$Hp=8$, the scatter is mainly along the $Hp-V_{T2}$ axis 
(see also Fig.~\ref{fig:err}).
 The relationship between $V\!-\!I_{C}$ and $Hp-V_{T2}$ 
cannot be represented by a single polynomial, hence we provide segments of 
calibration curves along with a color interval of their validity 
(Table~\ref{tab:pol}).  Within this interval, a Hipparcos $(V\!-\!I)_{H}$ is
\begin{equation}
(V\!-\!I)_{H}=\sum_{k=0}^4 c_k (Hp-V_{T2})^{k}. \label{eq:pol}
\end{equation}

\begin{table*}[htb]
\caption[]{\label{tab:pol}Polynomial transformation from $Hp-V_{T2}$ to
 $V\!-\!I_{C}$}
\begin{tabular}{ccccccc}\hline
Spectral Type & Color Range & $c_{0}$ & $c_{1}$ & $c_{2}$ & $c_{3}$ & 
$c_{4}$ \\ \hline
M,S &  $-0.20>Hp-V_{T2}\geq-0.80$ & $1.296$ & $-6.362$ & $-5.128$ & $-1.8096$ &
 $0.0$\phantom{0} \\ 
M,S &  $-0.80>Hp-V_{T2}\geq-2.50$ & $2.686$ & $-1.673$ & $0.0$ & $0.0$\phantom{0} & $0.0$\phantom{0} \\ 
C &  $-0.20>Hp-V_{T2}\geq-1.77$ & $1.297$ & $-4.757$ & $-4.587$ & $-2.4904$ &
$-0.5343$\\ 
C &  $-1.77>Hp-V_{T2}\geq-2.00$ & $3.913$ & $0.0$ & $0.0$ & $0.0$\phantom{0} & $0.0$\phantom{0} \\ 
\hline
\end{tabular}
\end{table*}

To calculate an epoch $(V\!-\!I)_H$, one should use the epoch $Hp$ photometry
and obtain $Hp-V_{T2}=b_0+b_1\times{Hp}$ (see Eq.~\ref{eq:hp}). Then, 
a polynomial
transformation given by Eq.~\ref{eq:pol} and Table~\ref{tab:pol}  
leads directly to the desired $(V\!-\!I)_H$ color index. However, 
there are numerous cases when it was not possible to determine
a slope $b_1$ in the $Hp-V_{T2}$ vs. $Hp$ plot, although the amplitude 
of $Hp$ variations indicated a likely change in $Hp-V_{T2}$ as well. 
Therefore, for all such stars with a light amplitude having the range 
between maximum and minimum luminosities,
$\Delta Hp>0.15$ (see entries H50-H49, ESA 1997, vol. 1), we
adopted the mean slope, i.e., the mean gradient given in Sect.~3.2.
A difficulty then is to find a point in the $Hp-V_{T2}$ vs. $Hp$ plot,
to which the mean slope can be applied in order to estimate an intercept $b_0$.
The median of the 3-5 brightest values of $Hp$ and the corresponding
median $Hp-V_{T2}$ color were adopted for such a `reference' point.

\begin{figure}[htb]
\resizebox{\hsize}{!}{\includegraphics{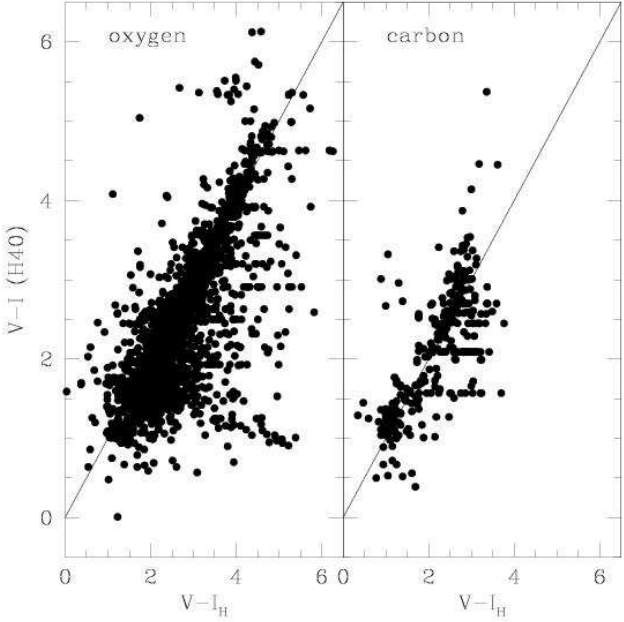}}
\caption[]{\label{fig:colcol} 
Hipparcos median $V\!-\!I$ \citep[ entry H40]{esa97} vs. newly derived
median $(V\!-\!I)_{H}$ in this study. The reasons for some very large
discrepancies are discussed in Sect.~\ref{sect:rem}.
}
\end{figure}

An important issue is to verify the system of our $(V\!-\!I)_H$ photometry
for red stars. The differences between the new median $(V\!-\!I)_H$ and 
the best available Hipparcos $V\!-\!I$ photometry (entry H40) are plotted in
Fig.~\ref{fig:colcol}. 
On average the two systems are consistent. The very red carbon stars are
an exception because their $(V\!-\!I)_H$ color indices reach 
saturation, whereas the Hipparcos $V\!-\!I$ index is not restrained. Then,
there are numerous cases where the newly derived $(V\!-\!I)_{H}$ values 
differ considerably from those in the Hipparcos Catalogue -- in extreme 
cases up to 3-4 mag. A closer look at these cases indicates various reasons
for such discrepancies. It could be duplicity, an incorrect target, severe 
extrapolation in color, etc. Noteworthy is the fact that the $I_{C}$ bandpass
given in \citet{esa97} is $\sim\!30$ nm wider on the red side than
the one published by \citet{bessell79}. Uncertainty in the location
of the red-side cutoff of the $I_{C}$-bandpass owing to different detectors
is known to be a major source of a small color-dependent bias 
($<\!0.1$ mag) in the ground-based photometry of red stars.

\subsubsection{Verification of the new $V\!-\!I$ color}

From the variety of available sources, we have chosen the two largest sets of 
ground-based Cousins $V\!-\!I$ data to test our $(V\!-\!I)_H$ color indices; 
that is \citet{koen02} for M stars and \citet{walker79} for carbon stars.   
We also selected the data of \citet{lahulla87}, which is an independent
source of $V\!-\!I$, albeit in the system of Johnson $V\!I$ which was not used 
in the calibration.

\begin{figure}[htb]
\resizebox{\hsize}{!}{\includegraphics{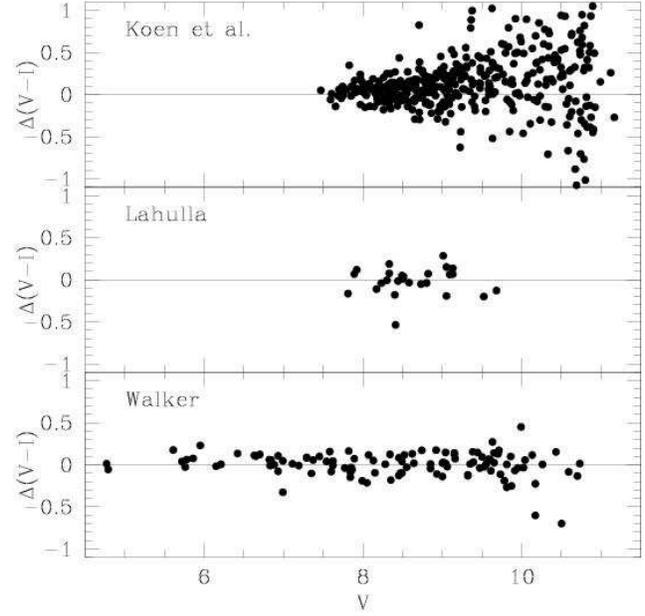}}
\caption[]{\label{fig:compar} 
Differences between our instantaneous $(V\!-\!I)_{H}$ color index and those of 
\citet{koen02,lahulla87,walker79}. The upper two panels represent
M stars, whereas the bottom panel contains carbon stars. The accuracy of 
our calibrated $V\!-\!I$ colors is clearly insufficient in the case of 
faint M dwarfs, which represent a large fraction of the \citet{koen02}
sample. 
}
\end{figure}
The differences, $(V\!-\!I)_{H}-(V\!-\!I)_{C}$, are plotted as a function of
ground-based $V$ (Fig.~\ref{fig:compar}). 
For the \citet{walker79} and \citet{lahulla87} datasets, the mean offset
$\langle(V\!-\!I)_{H}-(V\!-\!I)_{C}\rangle$ is not more than 
$+0.01$ mag; the scatter
of individual differences is 0.12 mag. The \citet{koen02} data 
are instrumental to test the reliability of $(V\!-\!I)_{H}$ for early-type
M stars, both dwarfs and giants. We note that at $V\!-\!I\approx2$ the
calibration curve is very steep (left panel, Fig.~\ref{fig:coltran}). 
At this $V\!-\!I$, a variation in $Hp-V_{T2}$ by only 0.01
mag corresponds to a 0.05 magnitude change in $V\!-\!I$. For relatively
bright Hipparcos stars $(V\!<\!9)$, the mean offset 
$\langle(V\!-\!I)_{H}-(V\!-\!I)_{C}\rangle$ 
is $+0.04$ but it increases to $+0.20$ for fainter stars $(9\!<\!V\!<\!11)$.
The scatter also rises from 0.13 to 0.40 in these two intervals.
A noticeable bias in the mean $(V\!-\!I)_{H}$ towards faint magnitudes
might be an indication of some residual systematic error either in
the Hipparcos $Hp$ epoch photometry or in Tycho-2 $V_{T2}$ magnitudes. 
As expected, rapidly increasing errors in $V_{T2}$ as a function of 
magnitude (Fig.~\ref{fig:err}) clearly set a limitation on
the accuracy of $(V\!-\!I)_{H}$.

\section{New $V\!-\!I$ and some applications}

We have calculated instantaneous (epoch) $(V\!-\!I)_H$ color indices for 
4414 M stars, 50 S stars 
from the list by \citet{vaneck98}, and 321 carbon stars, which include
R, N, and Hd sub-types. A condensed version of this effort is presented
in Table 7\footnote{
Table 7 is available only in electronic form at the Centre de Donn\'{e}es
Astronomiques de Strasbourg (CDS), France via anonymous ftp to 
\texttt{cdsarc.u-strasbg.fr (130.79.128.5)} or at
\texttt{http://cdsweb.u-strasbg.fr/cgi-bin/qcat?J/A+A/(vol)/(page)}},
which contains HIP number, GCVS name for variable
stars, median $Hp$ magnitude (entry H44, ESA 1997, vol. 1), 
5-to-95 percentile $Hp$ range or the $Hp$ `amplitude', coefficients
$b_0,b_1$ (if $b_1$ has not been determined, it is set equal to zero),
median $V\!-\!I$ from this study, spectral type (M, S, or C).

We note that about 2\% of Hipparcos M, S, and C stars do not have 
adequate Tycho-2 photometry and, hence, are not given in Table 7.
Those include some very bright stars and a number of faint stars.
More than a dozen stars of intermediate brightness with $8.0\!>\!Hp\!>\!5.0$
failed in the Tycho-2 photometry reductions due to poor astrometry,
high background and/or a parasitic signal, which corrupted the signal
from the target object.

\subsection{\label{sect:rem} Remarks on individual carbon stars}

We used the derived $(V\!-\!I)_H$ color index and in some cases individual
slopes from the $Hp-V_{T2}$ vs. $Hp$ plot to scrutinize the identity
of some Hipparcos carbon stars. If an anonymous field star is
measured instead of a real carbon star, it could yield a positive
slope in the fit of $Hp-V_{T2}$ vs. $Hp$. This is because the
$Hp$ measures have been overcorrected, using a $V\!-\!I$  
color index appropriate for an expected carbon star but not for the actual
target. On the other hand, the Tycho-2 $V_{T2}$ photometry appears to be 
insensitive to the color a star really has. The net result is
a very small or even positive slope. After identifying such cases,
we checked the 2MASS Atlas Images for the true location of a carbon
star in question. The offset in position is
given in Table~\ref{tab:off}. If a carbon star has incorrect coordinates
in \citet{alksnis01}, it is coded by `GCGCS:' in Remarks.
If an incorrect identification is already acknowledged in the Hipparcos
Catalogue, it is indicated by the `HIP note' in Remarks.   
In the case of contradictory spectral classifications, we list
only the alternative classification, since in nearly all such cases
Hipparcos spectral type is `R...'. 
None of them can be found in \citet{alksnis01};
therefore, the true identity of these stars has yet to be confirmed
by spectroscopic means. An exception is HIP 94049 = CGCS 4179 which is
a genuine carbon star (Houk, private communication; see also 
Table~\ref{tab:evans}).  

\begin{table}[htb]
\addtocounter{table}{+1}
\caption[]{\label{tab:off}Upper part: erroneous Hipparcos pointings of
carbon stars or a contradictory spectral classification with Hipparcos
indicating `R...' spectral type.
The offsets in coordinates, $\Delta$RA and $\Delta$Dec, are given in
($s$) and ($\arcsec$), respectively, in the sense `true position-Hipparcos'.
Lower part contains significant corrections `true position-GCGCS' 
required in \citet{alksnis01} to the positions of non-Hipparcos R stars 
from Table~\ref{tab:evans}.} 
\setlength{\tabcolsep}{1mm}
\begin{tabular}{rrrrl}\hline
HIP & CGCS & $\Delta$RA & $\Delta$Dec & Remarks \\ \hline
4266    &  &  & &  M0 (SAO) \\
14055    &  &  & &  M0 (SAO)\\
21392    &  &  & &  M0 (SAO) \\
22767   &  808 & $-21.0$ & $+9$ & HIP note \\ 
24548   &  893 & $0.0$ & $-242$ &  \\
29564    &  &  & & M0 (SAO) \\
29899   & 1226 & $+3.4$ & $+26$ & GCGCS: \\
35015   & 1615 & $+7.1$ & $-146$ & GCGCS:  \\
35119   & 1616 & $+0.3$ & $+59$ & HIP note, GCGCS: \\
37022   & 1787 & $-2.6$ & $+32$  & HIP note, GCGCS: \\
39337   & 2007 & $+16.7$ & $+31$ &  \\
40765   &  & & & G1V \citep{houk99} \\
44235   &  & & & not C-star? \citep{stephenson89} \\
75691   & 3614 & $+8.38$ & $+94$ & GCGCS: \\ 
83404   & 3762 & $-0.4$ & $-197$ & GCGCS:  \\
85148   & 3820 & $-1.6$ & $+58$ & GCGCS: \\
88170    &  &  & &  M0 (SAO) \\
94049    &  &  & &  C-star, not F5V \\
95024   & 4241 & $+5.3$ & $+10$ & HIP note, GCGCS: \\
106599  & 5371 & $-7.7$ & $+4$ & HIP note, GCGCS: \\
113840    &  &  & &  M0 (SAO) \\
118252  & 5970 & $-2.3$ & $-13$ & HIP note, GCGCS: \\ \hline
        & 258 & $-3.0$ & 0 & \\
        & 3765 & $-0.6$ & $-42$ & \\
        & 3810 & $+10.3$ & $+10$ & \\
        & 3813 & $+0.4$ & $+9$ & \\
        & 3864 & $+0.3$ & $-10$ & \\
        & 3939 & $+0.5$ & $-2$ & \\
        & 3966 & 0.0 & $+25$ & \\
        & 4042 & $+0.8$ & $-2$ & \\
        & 4168 & $+0.8$ & $-14$ & \\
        & 4498 & $+3.7$ & $-35$ & \\ 
\hline
\end{tabular}
\end{table}

\subsection{Duplicity and $V\!-\!I$ color index}\label{sect:dup}

\begin{figure}
\resizebox{\hsize}{!}{\includegraphics{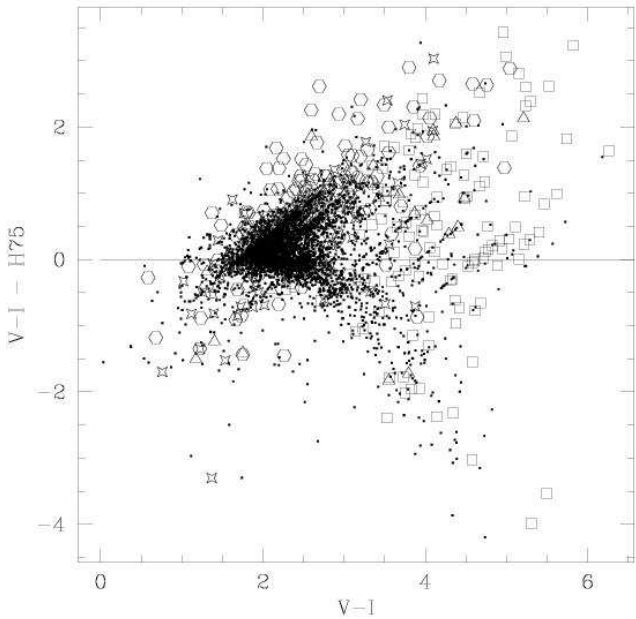}}
\caption[]{\label{fig:VI} Comparison of the $V\!-\!I$ color indices 
used in the preparation of the Hipparcos Catalogue (H75) with those 
rederived from a color transformation
based on $Hp - V_{T2}$. Only red stars with spectral types M and S 
are considered. Single-star solutions are depicted by black
dots, whereas the open symbols denote more complex solutions [hexagons --
component solutions C; triangles -- acceleration solutions G;
squares -- Variability-Induced Mover (VIM) solutions V; stars -- 
stochastic solutions X].
It is worth noting that nearly all datapoints in the upper right corner of
the diagram correspond to complex solutions, thus hinting at problems
encountered in the Hipparcos data processing for these stars.  
}
\end{figure}

\begin{figure}
\resizebox{\hsize}{!}{\includegraphics{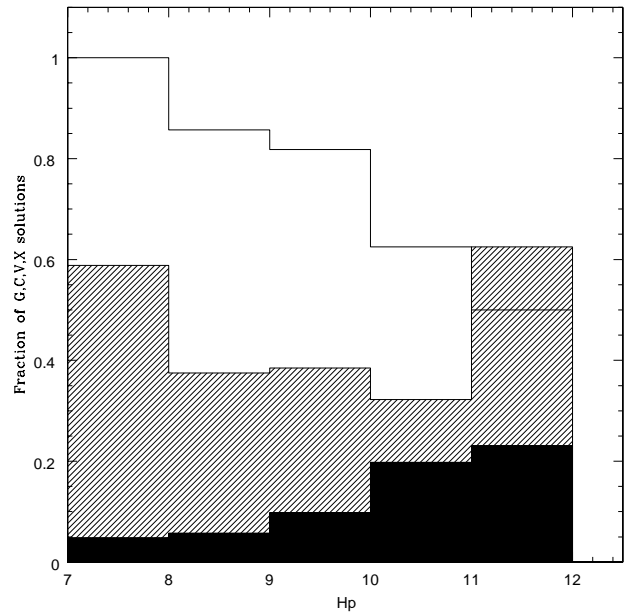}}
\caption[]{\label{fig:fracVI} A histogram showing the
fraction of red stars with a DMSA C,G,V or X solution in the Hipparcos
Catalogue (field H59), as a function of $Hp$ magnitude, for three different
sets of $\Delta(V-I)=(V\!-\!I)_H-(V\!-\!I)_{\rm H75}$, where $(V\!-\!I)_H$ is
the newly derived median color index and $(V\!-\!I)_{\rm H75}$ is Hipparcos 
median $V\!-\!I$ (entry H75). The unshaded histogram
shows the fraction of these DMSA solutions among the stars with
$\Delta(V\!-\!I)\!>2$; hatched for $2\ge\Delta(V\!-\!I)\!>1$; 
and dark-shaded for $1\ge\Delta(V\!-\!I)\!>\!-1$. 
The fraction of G,C,V, and X solutions clearly correlates with
$\Delta(V\!-\!I)$; except for the faintest stars which always have 
a large fraction of G,C,V,X solutions. 
}
\end{figure}

Perhaps, the star HIP 12086 = 15 Tri is a prototype of a very rare but
characteristic Hipparcos problem due to the neglected poor input 
coordinates. The declination of HIP 12086 listed in the Hipparcos
Input Catalogue \citep{esa92} is off
by $10\arcsec$, hence in the detector's instantaneous field of view
\citep[see][ vol. 3, Fig. 5.2]{esa97} the signal has apparently been
affected by the sensitivity attenuation profile. This kind of bias
is absent in the star mapper's instrumentation. As a result,
there is a very large positive slope in the $Hp-V_{T2}$ vs. $Hp$ plot.
Not only is the $Hp$ photometry clearly corrupted but the astrometry
is also degraded as indicated by unusually large errors in the astrometric
parameters. A similar effect of poor Hipparcos performance is known 
to be present, if the targets were wide binaries with separations in the
range $\sim\!15\arcsec-20\arcsec$ \citep{fabricius00a}. 
Here we list such binaries among red stars when the epoch $Hp$ photometry
is clearly biased: HIP 7762, 13714 \& 13716, 17750, 18465, 45343,
57473, 86961, 87820, 108943, 116191, 114994. We note that from
this list the revised astrometry is already available 
for  HIP 17750, 86961, 87820, 116191 \citep{fabricius00a}. 

Strictly speaking the $V\!-\!I$ index derived in this study for Hipparcos 
binary and multiple stars could be affected by the component(s) and,
hence should be considered with caution. On the other hand, a peculiar
$V\!-\!I$ value may very well signal a genuine problem,
be it of astrophysical or instrumental character. With this in
mind, we examined the location of complex astrometric solutions in the plot
given in Fig.~\ref{fig:colcol}. It turns out that certain areas, 
as seen in Fig.~\ref{fig:VI}, are heavily populated by such cases.
Why is it so? It is helpful to look at the relative fraction of 
DMSA C,G,V, and X solutions as a function of differences between our
median $(V\!-\!I)_H$ and Hipparcos $(V\!-\!I)_{\rm H75}$.  
Figure~\ref{fig:fracVI} shows that the relative fraction of
supposedly complex systems, i.e., binary or multiple stars, 
is abnormally high for red stars. For $\Delta(V\!-\!I)>1$ and $Hp<10$  
(see unshaded and hatched areas in Fig.~\ref{fig:fracVI}), the relative
fraction of such systems is 40\% and higher as compared to only 
$\sim\!10\%$ among the stars having correct $(V\!-\!I)_{\rm H75}$ index 
(dark-shaded histogram).  

Table~\ref{tab:spurious} lists all red stars with 
$(V\!-\!I)_H-(V\!-\!I)_{\rm H75}\!>\!2$.  As indicated from comparisons with 
an independent ground based $V\!-\!I$ color index (see column 3 in 
Table~\ref{tab:spurious}), such differences are real. In essence,
the stars listed in Table~\ref{tab:spurious} have been processed with  
the $(V\!-\!I)_{\rm H75}$ color index off by more than 2 mag! 
Among such stars, the fraction of DMSA C,G,V, and X solutions -- nearly
75\% -- is conspicuous in itself.
For example, in the case of HIP 19488 and HIP 91703, it is evident that 
speckle interferometry could not confirm duplicity and, hence the
Hipparcos DMSA/C solution must be spurious. This is nearly a watertight result
since the limiting angular resolution of speckle interferometry 
\citep{mason99,prieur02} is 2-3 times higher than the separation 
given in the Hipparcos Catalogue.  
The other stars with a DMSA/C solution listed in Table~\ref{tab:spurious} 
have not been observed so far under similar conditions nor are they listed 
in the Fourth Catalog of Interferometric Measurements of Binary Stars
\footnote{http://ad.usno.navy.mil/wds/int4.html}, 
so that their possible spurious nature has yet to be established. 
Nevertheless, the high fraction of failed confirmations of binarity for
Hipparcos stars with a DMSA/C solution 
\citep[e.g.,][]{mason99,mason01,prieur02} is indicative that many such
solutions might be spurious. We suspect that the phenomenon of such
non-existent binaries among the red stars could very well be rooted
in the improper chromaticity correction applied to these stars
due to the poor knowledge of their true $V\!-\!I$ color.

\begin{table*}
\caption[]{\label{tab:spurious}
All M, S spectral type stars 
with $\Delta(V\!-\!I)_H=(V\!-\!I)_H-(V\!-\!I)_{H75}>2$ (column 2),
where $(V\!-\!I)_H$ is a color index from this study.
When available, $\Delta(V\!-\!I)_{\rm 0}=(V\!-\!I)_H-(V\!-\!I)_{\rm obs}$, 
where $(V\!-\!I)_{\rm obs}$ is the ground-based photoelectric measurement.
The stars are ordered by increasing amplitude of variability $\Delta Hp$
(column 4).
The column labelled DMSA provides a type of Hipparcos solution assuming
more than one component. Angular separation between components, $\rho$,
is given for DMSA/C solutions only.
}

\begin{tabular}{rcclrccl}
\medskip\\
\hline
HIP &  $\Delta (V\!-\!I)_{H}$ &$\Delta(V\!-\!I)_{\rm 0} $ & $\Delta Hp$ & 
$Hp$ & DMSA & $\rho (\arcsec)$ & Remarks \\ 
\hline
 19488& 2.41 && 0.13& 9.535 & C & 0.18 & unresolved \citep{mason99}\cr
 78501& 2.19 && 0.14& 10.285& C & 0.17\cr
 24661& 2.31  && 0.15& 10.170& &\cr
 87221& 2.61  && 0.19& 8.763 & C & 0.17\cr
 87433& 2.27  & $-0.44$ & 0.28& 8.537 & & \cr
 76296& 2.26  && 0.33& 8.878 & C & 0.16\cr
 42068& 2.33  & $-0.49$ & 0.41& 8.511 & C & 0.18\cr
 91703& 2.65  && 0.46& 8.799 & C,V & 0.21 & unresolved \citep{prieur02}\cr
  7762& 2.03  && 0.48& 8.615 & X & & companion star at $20\arcsec$\cr
 84346& 2.05  && 0.61& 8.454 & V & & unresolved \citep{prieur02}\cr
100404& 2.14  & $-0.76$ & 0.61& 8.464 & V & & unresolved \citep{mason01}\cr
 37433& 2.17  && 0.64& 8.984 &\cr
 56533& 2.64  && 0.65& 8.581 & C & 0.24\cr
 84004& 2.40  && 0.78& 7.499 & X \cr
 80259& 2.19  && 0.91& 9.017 & V & & unresolved \citep{prieur02}\cr
 16328& 2.10  && 0.98& 9.612 & C & 0.30 \cr
% 35793& 2.13 0&& 1.32& 7.961 & G & & star embedded in nebula\cr
 90850& 2.32  && 1.38& 11.001&   \cr
 78872& 2.06  & $-0.47$ & 1.70& 9.841 & G \cr
   703& 2.43  & $-0.59$ & 2.09& 11.112& V \cr
  9767& 2.19  && 2.10& 9.773 & V \cr
 11093& 2.52  & \phantom{$-$}0.16 & 2.10& 9.756 & V \cr
% 91389& 3.03 3& $-1.60$ & 2.23& 7.283 & X & & close visual binary X Oph \cr
 89886& 3.27  && 2.26& 10.883& \cr
 96031& 2.29  && 2.64& 10.512& \cr
 75393& 3.06  & \phantom{$-$}0.32 & 2.73& 8.554 & V \cr
 16647& 3.23  && 2.87& 10.376& V \cr
 81026& 2.66  && 2.89& 11.538&   \cr
  1901& 2.61  & \phantom{$-$}1.27 & 2.97& 10.705& V & & unresolved \citep{prieur02}\cr
 86836& 3.43  && 3.15& 11.196& V \cr
 47066& 2.61  & \phantom{$-$}0.87 & 3.49& 10.073& V \cr
 57642& 2.32  & \phantom{$-$}0.78 & 3.60& 9.968 & V \cr
 60106& 2.04  && 3.81& 9.854 &   \cr
110451& 2.01  && 3.90& 11.460&   \cr
 94706& 2.81  & \phantom{$-$}0.67 & 3.97& 10.826& V & & T Sgr: composite spectrum \cr
 25412& 2.39  & \phantom{$-$}0.29 & 4.00& 9.974 & V \cr
 33824& 2.02  & \phantom{$-$}0.97 & 4.05& 9.922 &   & & \cr
\hline
\end{tabular}
\end{table*}

\subsection{Empirical effective temperatures of red stars}

Due to very complex spectra the red stars are cumbersome objects
for getting their effective temperature -- one of the fundamental
stellar parameters.  From different vantage points this has been
investigated, e.g., by \citet{bessell98,bergeat01,houdashelt00}.
Although the Cousins $V\!-\!I$ color index may not be the optimal
color to calibrate effective temperature due to the strong influence by
molecular absorption bands and possible reddening, nevertheless we
attempted to derive an empirical calibration of effective temperatures
for carbon and M giants. 
We used median $(V\!-\!I)_H$ for Hipparcos stars having interferometric
angular diameter measurements in $K$ ($\lambda=2.2$ $\mu$m) bandpass
\citep{dyck96,vanbelle97,vanbelle99} and corresponding effective
temperature estimates.  It is expected that the interstellar reddening
is low for the chosen Hipparcos stars because of their relative proximity
to the Sun.  In total, from these sources of effective temperature
determinations, we selected 27 small amplitude ($\Delta Hp\!<\!0.5$)  
M giants in the range $3.6>V\!-\!I>1.5$ and 16 carbon stars 
($3.8>V\!-\!I>2.4$) with no restriction on variability.
Similarly to \citet{dumm98} we adopted a linear relationship 
\begin{equation}
\log T_\mathrm{eff}=d_0+d_1(V\!-\!I).  \label{eq:teff}  
\end{equation}
For M giants, a least squares fit using Eq.~\ref{eq:teff} yields
$d_0=3.749\pm0.014$, $d_1=-0.087\pm0.007$, and the standard error
$\sigma_T=110$~K. For carbon stars the coefficients from the fit are:
$d_0=3.86\pm0.06$, $d_1=-0.153\pm0.021$, and the standard error
$\sigma_T=210$~K. Apparently, the effective temperature scale is
not satisfactory for carbon stars in terms of its precision. The color
mismatch between our median $(V\!-\!I)_H$ and the color of a variable star
at the time of interferometric observation can only partly explain
the noted large scatter. Another reason might include an unaccounted
for circumstellar extinction, carbon abundance and metallicity
effects on the color,
and rather large errors in the effective temperature determination.
The latter is discussed in detail by \citet{dyck96}. 
An alternative scale of effective temperatures for carbon stars
is given by \citet{bergeat01}, although it may have the same kind
of inherent problems.
We note that the slope $d_1$ for M giants is 2.5 times larger than in 
\citet{dumm98}. The main reason for that is a stretched color 
scale of Hipparcos  $V\!-\!I$ (see Fig.~\ref{fig:colcol}).  It
is felt that the empirical effective temperature scale based 
on $V\!-\!I$ color has a limited use, in particular for carbon stars.
Near infrared observations in $JHKL$ bandpasses should be used
to obtain better estimates of effective temperature for the coolest
stars.  With the advent of large optical interferometers the number of
precise angular diameters for cool and red stars undoubtedly  
will increase substantially. However, an equal effort should be  
invested in deriving reliable bolometric fluxes, which are equally
important in establishing a precise scale of effective temperatures.

\section{Summary and conclusions}

The main result of this work is demonstrating the feasibility of the 
$Hp-V_{T2}$ color index in studies of red stars. This color
index is tightly correlated with the Cousins $V\!-\!I$
color and, thus, allows us to derive an independent estimate of
$(V\!-\!I)_H$ for carbon, M and S stars. Such estimates are indispensable
in the analysis of red variable stars, which have been little studied in 
the Cousins $V\!I$ system.

We have shown that a considerable fraction of Hipparcos best estimates of
$V\!-\!I$ color index (entry H40, ESA 1997) for red stars might be in error 
by more than a full magnitude. Conspicuously, among the most
discrepant cases we find an unusually large number of DMSA C,G,V, and X
solutions implicating a binary or multiple star status for these stars.
On the other hand, extensive speckle interferometric observations have 
largely failed to confirm the binarity, despite the 2-3 times better 
angular resolution.
This strongly suggests that some DMSA C,G,V,X solutions are not real
and maybe due to the poor knowledge of the $V\!-\!I$ color index, which 
served as a measure of star's color in both photometric and astrometric
reductions by the Hipparcos consortia. 

However, our attempts have not succeeded in improving the astrometry
for single red stars. It was expected that an incomplete correction for
the chromaticity effects should leave a color-related `jitter' in the
abscissa data at the level of 1-3 mas due to incorrect $V\!-\!I$, used in 
accounting for these effects. Surprisingly, we were not able to find
clear traces of residual chromaticity effects, for instance,
in carbon star Hipparcos astrometry. Either they have
been somehow accounted for in the original Hipparcos reductions or
they are insignificant.

On the other hand, the re-analysis of so-called Variability-Induced
Movers (VIM) has benefited substantially from the new set of $(V\!-\!I)_H$
color indices. As indicated in Sect.~\ref{sect:dup}, some of the DMSA/V
solutions are suspected to be not warranted. Much finer analysis of
all DMSA/V solutions for red  stars \citep{pourbaix02}  
provides strong evidence that nearly half of DMSA/V solutions are 
not justified, mainly thanks to reliable $V\!-\!I$ colors now
available at all phases of lightcurve for long-period variables such 
as Miras. This knowledge of $V\!-\!I$ colors could be useful to further
investigate other difficult systems having an extreme and changing color 
in combination with hints of duplicity, which can be resolved with 
interferometric means.

\begin{acknowledgements}

We thank A. Alksnis, C. Barnbaum, C. Fabricius, N. Houk, U. J\o rgensen, 
D. Kilkenny, B. Mason, J. Percy, and G. Wallerstein for their expert advice 
and help at various stages of this project. This work is supported in part by 
the ESA Prodex grant C15152/NL/SFe(IC). 
I. P. would like to thank the staff of IAA for their generous hospitality
during his stay in Brussels. L. N. B. thanks the staff of
Siding Spring Observatory for hospitality and technical support. 
T. L. E.  thanks his former colleagues at SAAO, especially R. M. Banfield 
and A. A. van der Wielen, for their assistance.
T. L. has been supported by the Austrian Academy of Science
(APART programme). The work with the Vienna APT has been made possible
by the Austrian Science Fund under project numbers P14365-PHY and
S7301-AST.  J. S. acknowledges a travel 
support from the National Science Foundation grant AST 98-19777 to USRA.
Illuminating comments and a number of suggestions by the referee, 
M. Bessell, are also greatly appreciated.
This research has made use of the SIMBAD database operated at CDS, Strasbourg,
France. This publication makes use of data products from the Two Micron All
Sky Survey, which is a joint project of the University of Massachusetts
and the Infrared Processing and Analysis Center/California Institute of
Technology, funded by the National Aeronautics and Space
Administration and the National Science Foundation.
\end{acknowledgements}

\bibliographystyle{aa}
\bibliography{biblio}
\end{document}